\documentclass[a4paper,fleqn]{cas-sc}

\usepackage{placeins}
\usepackage[numbers]{natbib}

\begin{document}
\let\WriteBookmarks\relax
\def\floatpagepagefraction{1}
\def\textpagefraction{.001}
\shorttitle{Westcott $g$ Factors Extended to Arbitrary Neutron Energy Spectra}
\shortauthors{D.A. Matters et~al.}

\title[mode = title]{Westcott $g$ Factors Extended to Arbitrary Neutron Energy Spectra}      
\cortext[1]{Corresponding author}

\author[1]{D.A. Matters}[type=editor,
                        orcid=0000-0003-4288-2817]
\cormark[1]
\ead{damatters@lbl.gov}
\ead[url]{https://github.com/DMatters}
\affiliation[1]{organization={Nuclear Science Division, Lawrence Berkeley National Laboratory},
	city={Berkeley},
	state={California},
	statesep={}, 
	postcode={94720},
	country={USA}}

\author[2]{A.M. Hurst}
\affiliation[2]{organization={Department of Nuclear Engineering, University of California at Berkeley},
	city={Berkeley},
    state={California},
	statesep={}, 
	postcode={94720},
	country={USA}}

\author[3]{T. Kawano}
\affiliation[3]{organization={Theoretical Division, Los Alamos National Laboratory},
    city={Los Alamos},
    state={New Mexico},
	statesep={}, 
	postcode={87545},
	country={USA}}

\begin{abstract}
Westcott $g$ factors are used in Neutron Activation Analysis (NAA) and Prompt Gamma-ray Activation Analysis (PGAA) to evaluate the impact of non-$1/v$ behavior in the neutron-capture cross sections of certain nuclei on activation product yields. This non-$1/v$ behavior arises from the presence of neutron resonances in the neutron-capture cross sections that overlap with the source neutron spectrum at low ($<5$~eV) energies. Historically, Westcott $g$ factors that have been cataloged for NAA and PGAA applications are the result of calculations that assume a Maxwellian neutron flux distribution with a given temperature. In this work, we use this approach with updated neutron-capture cross sections from the Evaluated Nuclear Data File, version VIII.1 (ENDF/B-VIII.1) to tabulate Westcott $g$ factor values for a broad range of Maxwellian distribution temperatures, comparing the results against currently-available $g$ factors from International Atomic Energy Agency tables and other sources. It was discovered during this analysis that the use of guided thermal and cold-neutron beams at certain facilities necessitates an approach for evaluating Westcott $g$ factors based on arbitrary non-Maxwellian spectra. In this paper, we present an approach for calculating $g$ factors with user-specified neutron spectra, and we demonstrate these methods to obtain Westcott $g$-factors for guided- and cold-neutron beams at the Budapest Research Reactor and the Forschungsreaktor M{\"u}nchen II reactor. As part of this work, open-source software has been developed that can be used to perform these calculations for applications in PGAA and NAA experiments.
\end{abstract}

\begin{keywords}
Neutron activation analysis \sep Neutron-capture reactions \sep Evaluated nuclear data libraries \sep Westcott $g$ factors \sep Guided neutron beams
\end{keywords}

\maketitle

\section{Introduction\protect\label{sec:Introduction}}
Below neutron energies of approximately 5~eV, the neutron-capture cross sections for most nuclei are inversely proportional to the velocity $v$ of the incident neutron. This relationship is sometimes known as the $1/v$ law, although there are a number of notable exceptions to this behavior, termed non-$1/v$ absorbers or irregular nuclei~\citep{IAEA,Mughabghab18,Molnar}. These irregular nuclei have neutron-capture resonances that overlap the thermal-energy range, which distorts the usual $1/v$ behavior of the capture cross section at thermal and lower velocities.

An effective neutron-capture cross section $\sigma_{\text{eff}}$ is defined in terms of the neutron-capture cross section $\sigma(v)$ and the normalized neutron flux distribution $\phi(v)$ by the equation
\begin{equation}
	\sigma_{\text{eff}} = \frac{1}{v_0} \frac{\int_{0}^{\infty}\sigma(v) \phi(v)dv}{\int_{0}^{\infty} \phi(v)/vdv}\label{eq:effective_cs},
\end{equation}
where $v_{0}=2200$~m/s is the thermal-neutron velocity~\citep{Westcott55, Westcott60}. In Neutron Activation Analysis (NAA) and Prompt Gamma-Ray Activation Analysis (PGAA) involving thermal and cold neutron beams, a thermal-equivalent neutron flux $\Phi_0$ can be defined by integrating over the distribution $\phi(v)$:
\begin{equation}
	\Phi_0 = \Phi \int_{0}^{\infty}\frac{v_0}{v}\phi(v)dv\label{eq:thermal_equivalent_flux}.
\end{equation}
Here, $\Phi$ is the actual number of neutrons that reach a unit surface area of the target per second~\citep{Molnar}. The measured quantity in NAA and PGAA is the reaction rate $R$~\citep{Byrne}, which for thin targets with $N$ nuclei in the neutron beam is:
\begin{equation}
	R = N \sigma_{\text{eff}} \Phi_0\label{eq:reaction_rate}.
\end{equation}
In practice, Eq.~\ref{eq:reaction_rate} must be adjusted to account for neutron absorption in the target mass, and any contributions from epithermal neutrons in the spectrum. These corrections are addressed in detail in Refs.~\citep{Molnar, Hurst15}.

To account for the non-$1/v$ behavior of irregular nuclei in a simple way for applications in NAA or PGAA, a correction factor known as the Westcott $g$ factor was devised~\citep{Westcott55,Westcott60}. For example, experiments can involve measurements that use comparator isotopes with well-known neutron-capture cross sections such as $^{35}$Cl and $^{37}$Cl to normalize $\gamma$-ray intensities for isotopes of interest~\citep{Matters16, Hurst19}. These comparator isotopes are generally $1/v$ absorbers, while the isotope of interest may not be, necessitating adjustment using a Westcott $g$ factor~\citep{Choi14}. The Westcott $g$ factor is an integral quantity, defined so that for a thermal-neutron source,
\begin{equation}
	\sigma_{\text{eff}} = g \sigma_0 \label{eq:g_factor},
\end{equation}
where $\sigma_{0}$ is the neutron-capture cross section for thermal ($v_{0}=2200$~m/s) neutrons.  Using the definition of the effective cross section from Eq.~\ref{eq:effective_cs}, the Westcott $g$ factor becomes
\begin{equation}
	g =\frac{\sigma_{\text{eff}}}{\sigma_0} = \frac{1}{\sigma_{0} v_{0}} \frac{\int_{0}^{\infty}\sigma(v) \phi(v)dv}{\int_{0}^{\infty}\phi(v) / v dv}\label{eq:gfactor_integral_velocity}.
\end{equation}
For a regular nuclide that follows the so-called $1/v$ law, the expression in Eq.~\ref{eq:gfactor_integral_velocity} evaluates to $g=1$ for any thermal-neutron flux distribution $\phi(v)$.  Equivalent forms for Eq.~\ref{eq:gfactor_integral_velocity}, written in terms of the neutron energy $E$ and wavelength $\lambda$, can be derived by applying the appropriate variable substitution, where $E=\frac{1}{2} m_n v^2$ and $\lambda = \frac{h}{m_n v}$ ($h$ is Planck's constant and $m_n$ is the neutron mass). In these expressions, the thermal-neutron energy and wavelength are $E_{0}=0.025$~eV and $\lambda_0 = 1.8$~{\AA}, respectively:
\begin{equation}
	g =\frac{1}{\sigma_{0} \sqrt{E_{0}}} \frac{\int_{0}^{\infty}\sigma(E) \phi(E)dE}{\int_{0}^{\infty}\phi(E) / \sqrt{E} dE}\label{eq:gfactor_integral_energy}
\end{equation}
and
\begin{equation}
	g =\frac{\lambda_0}{\sigma_{0}} \frac{\int_{0}^{\infty}\sigma(\lambda) \phi(\lambda)d\lambda}{\int_{0}^{\infty}\phi(\lambda) \lambda d\lambda}\label{eq:gfactor_integral_wavelength}.
\end{equation}

Westcott $g$ factors are often calculated assuming a Maxwellian-distributed neutron spectrum, and in this convention they are typically written with a subscript $T$, where $g_T$ is the $g$ factor for a neutron spectrum defined as a Maxwellian distribution with temperature $T$~\citep{IAEA, Pritychenko25}.  In this work we present a methodology for calculating Westcott $g$ factors for arbitrary neutron spectra without a clearly-defined temperature, so the subscript $T$ will be suppressed unless we are expressly referring to a Maxwellian-spectrum $g$ factor.

In the sections that follow, we present two alternative methods for calculating Westcott $g$ factors, extending these methods from Maxwellian distributions to arbitrary neutron spectra, such as those measured experimentally at PGAA and NAA facilities. These methods are implemented in open-source software tools that can be employed by users to calculate $g$ factors for any target nucleus with available neutron-capture cross section data. Results from using these software tools are then compared to show the advantages of using cross sections from evaluated libraries and experimental neutron energy spectra instead of approximations to calculate $g$ factors.

\section{Calculating Westcott $g$ Factors\protect\label{sec:Calculations}}
It can be useful to define a neutron density function $p(v)$, which is the distribution of neutrons with a velocity $v$ in a unit volume. The neutron density is related to the flux distribution $\phi(v)$ (the distribution of neutrons crossing a unit surface per second) according to the relation $p(v) = \phi(v)/v$, and it is normalized such that $\int_{0}^{\infty} p(v) dv = 1$. Written in terms of $p(v)$, the expression for the Westcott $g$ factor from Eq.~\ref{eq:gfactor_integral_velocity} becomes 
\begin{equation}
	g=\frac{\int_{0}^{\infty}\sigma(v) v p(v)dv}{\sigma_{0} v_{0}}\label{eq:gw_integral}.
\end{equation}
Values for the thermal-neutron capture cross sections $\sigma_{0}$ are needed in order to evaluate the expression in Eq.~\ref{eq:gw_integral}. These are readily available from sources such as the \textit{Atlas of Neutron Resonances}~\citep{Mughabghab18}, IAEA tables of pile oscillator results~\citep{Firestone21}, the IRDFF-II neutron metrology library~\citep{Trkov20}, and the Evaluated Gamma-ray Activation File (EGAF)~\citep{EGAF}. The recently-developed pyEGAF software package~\citep{Hurst23,pyEGAF-pypi} allows users to access, manipulate, and analyze neutron-capture $\gamma$-ray data in EGAF database, including querying the database for $\sigma_{0}$ values.  

However, evaluating the integral in Eq.~\ref{eq:gw_integral} directly can be difficult, depending on the availability and quality of cross section data $\sigma(v)$ at various neutron velocities other than $v_0$. Experimental cross-section data are available at discrete energies, and libraries such as the Evaluated Nuclear Data File (ENDF)~\citep{Brown18} include modeled cross sections where experimental data are unavailable or insufficient.

\subsection{Irregularity Functions\protect\label{subsec:Irreg_funcs}}
In order to avoid integrating over the cross section $\sigma(v)$, an approximate method for calculating Westcott $g$ factors is outlined in Ref.~\citep{Molnar} that involves integrating over low-energy neutron resonances which give rise to the non-$1/v$ behavior in the capture cross section. Neutron resonance data from the \textit{Atlas of Neutron Resonances}~\citep{Mughabghab18} are used to define an irregularity function $\delta_{0}(v)$, which is the Lorentzian part of the Breit-Wigner formula~\citep{Molnar,Byrne}. The irregularity function describes the behavior of the cross section as a function of neutron velocity near a resonance of energy $E_{r}$ and total width $\Gamma$, according to the function
\begin{equation}
	\delta_{0}(v)=\frac{\left(E_{r}-\frac{1}{2}m_{n}v_{0}^{2}\right)^{2}+\Gamma^{2}/4}{\left(E_{r}-\frac{1}{2}m_{n}v^{2}\right)^{2}+\Gamma^{2}/4}.\label{eq:lorentzian_single}
\end{equation}  
The cross section $\sigma(v)$ can be approximated at any neutron velocity using the irregularity function $\delta_0(v)$~\citep{Molnar}:
\begin{equation}
	\sigma(v) \approx \frac{\sigma_0 v_0}{v} \delta_0(v) \label{eq:irregularity_cs}.
\end{equation}
Substituting Eq.~\ref{eq:irregularity_cs} into Eq.~\ref{eq:gw_integral} results in an approximation for the  Westcott $g$ factor:
\begin{equation}
	g \approx \intop_{0}^{\infty}\delta_{0}(v) p(v)dv\label{eq:gw_integral_irregularity}.
\end{equation}
When there is more than one resonance, the lowest-energy resonance is commonly used to define the function $\delta_{0}(v)$ in Eq.~\ref{eq:lorentzian_single}~\citep{Molnar}. However, doing so implicitly assumes that the lowest-energy resonance provides the dominant contribution to the non-$1/v$ behavior of the neutron-capture cross section. In general, this may not always be true, and the approximation inherent in Eq.~\ref{eq:gw_integral_irregularity} may not be valid in cases where there are multiple low-energy resonances or resonances reported with negative energies relative to the neutron-separation energy~\citep{Mughabghab18}.  In these cases, a more accurate method for calculating Westcott $g$ factors is to integrate over the entire neutron-capture cross section as defined in reaction data libraries such as the ENDF.

Resonance parameters, including the total widths $\Gamma$ and energies $E_{r}$, are tabulated in ENDF/B-VIII.1~\citep{Brown18, Nobre24}, the \textit{Atlas of Neutron Resonances}~\citep{Mughabghab18}, and the \textit{Handbook of Prompt Gamma Activation Analysis}~\citep{Molnar}. The resonance parameters from these sources were used to evaluate Westcott $g$ factors using the irregularity function method outlined above for six representative irregular nuclei: $^{83}$Kr, $^{115}$In, $^{149}$Sm, $^{151}$Eu, $^{157}$Gd, and $^{176}$Lu. These six non-$1/v$ absorbers provide good exemplars for demonstrating the calculation of Westcott $g$ factors, with cross sections that range from nearly-$1/v$ behavior (e.g., $^{83}$Kr) to highly irregular (e.g., $^{176}$Lu), involving a range of resonance energies and widths, as listed in Table~\ref{Table:Resonance_parameters}.

\begin{table}[pos=h!]
	\centering{}\caption{Neutron-capture cross section resonance energies $E_{r}$ and total widths $\Gamma$ for select non-$1/v$ nuclei, for low-energy resonances with $E_{r} < 5$~eV, from ENDF/B-VIII.1~\citep{Brown18,Nobre24}, the \textit{Handbook of Prompt Gamma Activation Analysis} (PGAA Handbook~\citep{Molnar}), and the \textit{Atlas of Neutron Resonances}~\citep{Mughabghab18}. Bold text indicates the parameters used to calculate Westcott $g$ factors via the irregularity function method in this work. \protect\label{Table:Resonance_parameters}}
	\begin{tabular}{|p{1cm}||p{2cm}p{2cm}|p{2cm}p{2cm}|p{2cm}p{2cm}|}
		\hline
		\ & \multicolumn{2}{c|}{ENDF/B-VIII.1$^{a}$} & \multicolumn{2}{c|}{\textit{Atlas of Neutron Resonances}} & \multicolumn{2}{c|}{PGAA Handbook}\\
		\hline
		Isotope & $E_{r}$ (eV) & $\Gamma$ (meV) & $E_{r}$ (eV) & $\Gamma$ (meV) & $E_{r}$ (eV) & $\Gamma$ (meV) \\
		\hline 
		\hline
		$^{83}$Kr & \textbf{-9.81} & \textbf{384} & \textbf{-9.81} & \textbf{252} & -3.9 & 245  \\
		\hline 
		$^{115}$In & \textbf{1.457} & \textbf{75.0} & \textbf{1.457} & \textbf{72} & 1.457 & 75.04  \\
		& 3.85 & 81.4 & 3.85 & 81 & 3.86 & 81.2  \\
		\hline 
		$^{149}$Sm & -1.127 & 71.9 & -1.127 & 64.9 & -1.5 & 87.4  \\
		& \textbf{0.0973} & \textbf{63.4} & \textbf{0.0973} & \textbf{62.9} & 0.0973 & 61  \\
		& 0.872 & 61.5 & 0.872 & 60.8 & 0.872 & 60.5  \\
		& 4.94 & 65.9 & 4.94 & 64 & 4.95 & 63.1  \\
		\hline 
		$^{151}$Eu & \textbf{-0.0609} & \textbf{105.3} & \textbf{-0.00362} & \textbf{95.8} & 0.00361$^{b}$ & 95.8 \\
		& 0.321 & 79.6 & 0.327 & 79.5& 0.321 & 79.6 \\
		& 0.46 & 87.7 & 0.465 & 87 & 0.461 & 87.7 \\
		& 1.055 & 88.2 & 1.054 & 85.7 & 1.055 & 85.2 \\
		& 1.815 & 91.0 & 1.806 & 95 & 1.83 & 90 \\
		& 2.717 & 94.2 & 2.717 & 92 & 2.717 & 93.2 \\
		& 3.368 & 95.2 & 3.370 & 88 & 3.366 & 95.2 \\
		& 3.71 & 93.9 & 3.711 & 92 & 3.71 & 94 \\
		& 4.79 & 91.2 & 4.796 & 66 & 4.78 & 90 \\
		\hline 
		$^{157}$Gd & \textbf{0.0314} & \textbf{107.7}$^{c}$ & \textbf{0.0314} & \textbf{107} & 0.0314 & 106 \\
		& 2.825 & 97.3 & 2.827 & 98 & 2.83 & 97 \\
		\hline 
		$^{176}$Lu & \textbf{0.1413} & \textbf{62.4} & \textbf{0.1379} & \textbf{65.0} & 0.141 & 62.4  \\
		& 1.565 & 59.5 & 1.569 & 59 & 1.565 & 59.5  \\
		& 4.36 & 68.4 & 4.316 & 65 & 4.36$^{d}$ & 68.4$^{d}$  \\
		\hline 
	\end{tabular}
	$^{a}$Resonance parameters were taken from the Breit-Wigner tables, except where otherwise noted.\\
	$^{b}$A negative sign, omitted in Ref.~\citep{Molnar}, is present in ENDF/B-VI.8 (the source for parameters in the PGAA Handbook).\\
	$^{c}$The total resonance width is equal to the sum of the Reich-Moore resonance widths listed in ENDF/B-VIII.1.\\
	$^{d}$Resonance parameters evidently used to calculate the $^{176}$Lu $g$-factors in the PGAA Handbook; see Table~\ref{Table:Irregularity_comparison}.
\end{table}

\subsection{Cross Sections from ENDF/B-VIII.1\protect\label{subsec:CS_integration}}
\begin{figure}[pos=h!]
	\centering{}\includegraphics[width=0.9\textwidth]{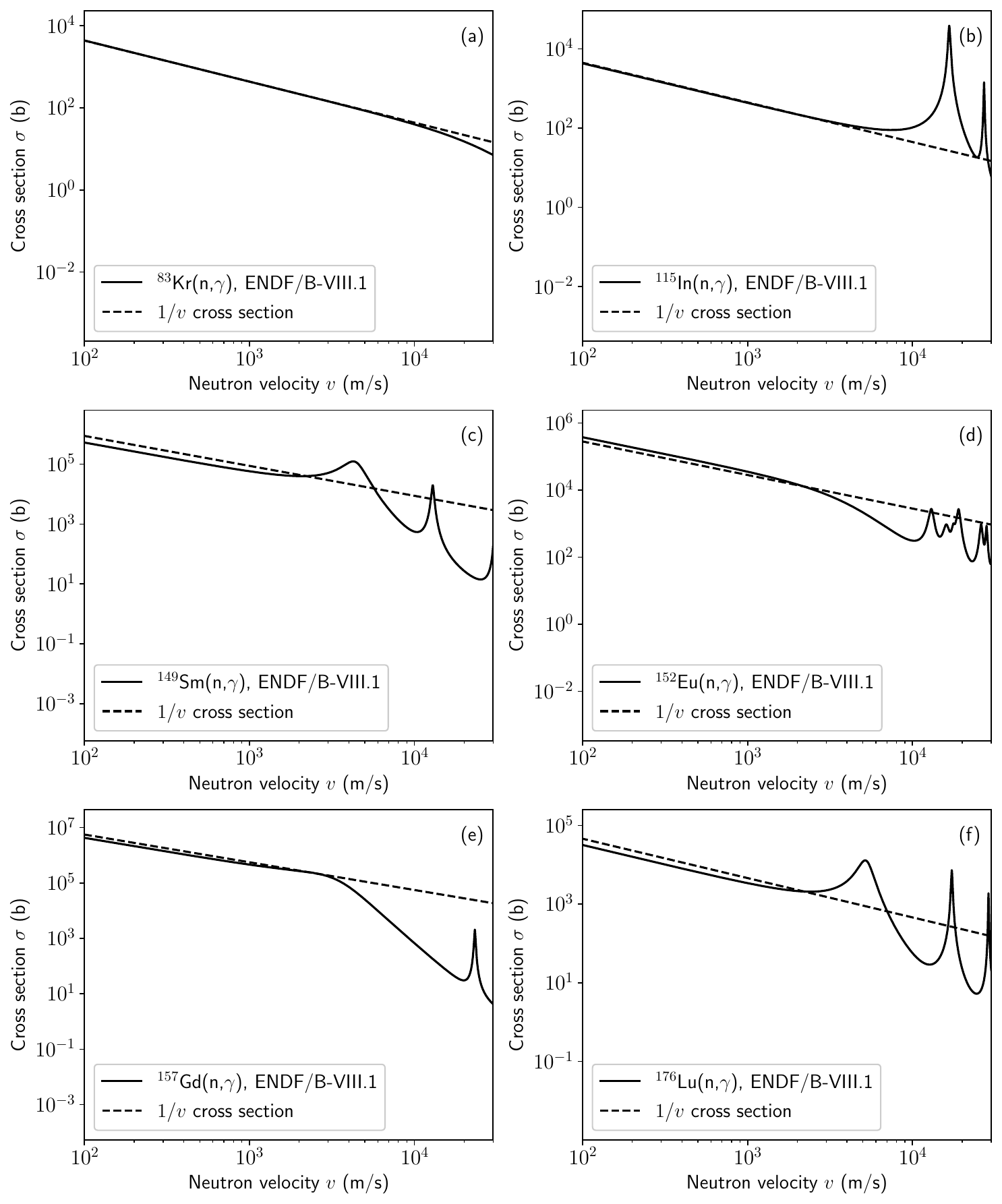}\caption{Low-energy ($E \leq 5$~eV) neutron-capture cross sections from ENDF/B-VIII.1~\citep{Nobre24} for each of the isotopes listed in Table~\ref{Table:Resonance_parameters}. The ENDF/B-VIII.1 cross sections ($\sigma(v)$) are plotted against hypothetical linear (on the log-scale plot) $1/v$ cross sections to show the irregular behavior of these nuclei due to the presence of capture resonances. \protect\label{fig:Cross_sections}}
\end{figure}

While irregularity functions provide a straightforward method for calculating Westcott $g$ factors from  neutron-capture resonance widths such as those in Ref.~\citep{Mughabghab18}, the method can produce errors for nuclei that have multiple low-energy resonances. To illustrate the effect of low-energy resonances, neutron-capture cross sections for each of the nuclei listed in Table~\ref{Table:Resonance_parameters} are shown in Fig.~\ref{fig:Cross_sections}.

It is evident from Fig.~\ref{fig:Cross_sections}(b-f) that for a number of irregular nuclei, the capture cross section below 5~eV (\mbox{$v < 3 \times 10^{4}$~m/s}) cannot be approximated simply with a single Lorentzian lineshape. A more appropriate method that can be applied generally to evaluate $g$ is to use the full neutron-capture cross section defined in evaluated nuclear data libraries such as ENDF/B-VIII.1 and numerically integrate Eq.~\ref{eq:gw_integral} directly. To do this requires defining the cross section $\sigma(v)$ at all neutron velocities.

In practice, this was done in two different ways. First, the \texttt{DeCE} (Descriptive Correction of ENDF-6 format) tool~\citep{Kawano19, DeCE-GitHub}, a C++ code that interfaces with cross sections in the ENDF-6 format, was modified to calculate Westcott $g$ factors by interpolating the cross section over a range of energies and numerically integrating Eq.~\ref{eq:gfactor_integral_energy}. Second, a purpose-built tool called \texttt{WestcottFactors} was written in Python to ingest ENDF cross-section data sourced from the Generalized Nuclear Database Structure (GNDS) hierarchies~\citep{Brown20}, interpolate $\sigma(v)$, and perform the integration in Eq.~\ref{eq:gw_integral}. Both the \texttt{DeCE} and \texttt{WestcottFactors} codes were then modified with the functionality to ingest CSV-formatted experimental neutron-flux distributions, to provide greater functionality to users. These codes are described in further detail below, and results are compared in Sec.~\ref{sec:Results} to demonstrate consistency between the two methods.  The advantage of providing two different tools for calculating Westcott $g$ factors with Maxwellian or user-defined neutron spectra are that users can choose to use either tool based on their comfort with different ENDF formats. Traditional users of ENDF already familiar with the ENDF-6 format may find the \texttt{DeCE} method appealing, while early adopters of the GNDS format may elect to use the \texttt{WestcottFactors} toolkit. \texttt{WestcottFactors} is also a stand-alone code packaged with cross-section data extracted from ENDF/B-VIII.1 in publicly-accessible GitHub~\citep{WestcottFactors-GitHub} and Python Package Index (PyPI) repositories~\citep{WestcottFactors-PyPI}, and the only requirement for calculating Westcott $g$ factors is a cursory knowledge of Python.

\subsubsection{\texttt{\textup{DeCE}}} The neutron-capture cross section data can be reconstructed from the resonance parameters in ENDF/B-VIII.1, and they are combined with the cross section data in MF (quantity) number 3, MT (reaction type) number 102 in the ENDF-6 formatted file for each nuclide in the library. These combined data are again stored in the MF=3 and MT=102 section for neutron energies ranging from 0.01~meV to 20~MeV. Codes such as \texttt{DeCE}~\citep{Kawano19} can be used to interface with data in the ENDF library using the legacy ENDF-6 plain-text format, and extract or manipulate the data for the purpose of evaluating Westcott $g$ factors. A GitHub fork to the development branch of \texttt{DeCE}~\citep{DeCE_fork_develop-GitHub} was modified to include routines for calculating Westcott $g$ factors for Maxwellian neutron flux distributions using cross-section data in the ENDF-6 format~\citep{Brown23}.

\subsubsection{\texttt{\textup{WestcottFactors}}} Starting with the ENDF/B-VIII version of the library, nuclear data were released in the GNDS format~\citep{Brown20} alongside the ENDF-6 format. The GNDS format was developed to make nuclear data more accessible through the use of modern computational tools, and it was written in an XML-hierarchichal format that lends itself to interpretation and manipulation by a variety of Python data science packages such as \texttt{pandas}~\citep{McKinney10}. In this work, we developed Python scripts to work with \texttt{FUDGE}~\citep{Mattoon23, FUDGE-GitHub} to extract data from ENDF/B-VIII.1 and interpolate the neutron-capture cross sections stored in MF=3, MT=102. As part of the \texttt{WestcottFactors} package, a representative Jupyter notebook is included that provides users with the capability to extract cross sections from GNDS-formatted data, although this is not necessary to use the tool because capture cross-section data from ENDF/B-VIII.1 are already included with the package~\citep{WestcottFactors-GitHub, WestcottFactors-PyPI}.

Numerical integration methods in the \texttt{scipy} package~\citep{Virtanen20} were then used to integrate over the cross sections and neutron flux distributions to evaluate Westcott $g$ factors according to Eq.~\ref{eq:gw_integral}. The scripts used to perform these calculations have been packaged into the \texttt{WestcottFactors} tool, along with CSV-formatted cross-section data parsed from ENDF/B-VIII.1 and explanatory Jupyter notebooks to demonstrate and interact with the code. The open-source \texttt{WestcottFactors} package is available via a publicly-accessible GitHub repository and PyPI package~\citep{WestcottFactors-GitHub, WestcottFactors-PyPI}. \texttt{WestcottFactors} leverages modern scientific computing libraries to provide a robust method for calculating Westcott $g$ factors in a way that is transparent and traceable end-to-end. It can also be easily modified to ingest data from future versions of the ENDF library as they are released.

\subsection{Neutron Flux Distributions\protect\label{subsec:Neutron_dist}}
Neutron-activation experiments involving irregular nuclei can require adjustments to the measured activities using Westcott $g$ factors. However, there are a variety of neutron sources used in experiments, each with unique energy spectra determined by their origins, moderators, and beamlines. For example, PGAA and NAA experiments are commonly performed at facilities that incorporate research reactors with cryogenic moderators and guided neutron beamlines, such as the Budapest Research Reactor (BRR) and the Forschungsreaktor M{\"u}nchen II (FRM~II) reactor.

Several references list Westcott $g$ factors for Maxwellian neutron flux distributions~\citep{IAEA, Molnar, Nichols08, vanSluijs15, Holden99, Pritychenko25, Chand25}, as do the original papers by Westcott on this topic~\citep{Westcott55, Westcott60}. Other simplified distributions, such as those for idealized guided-neutron beams, are also included in some references~\citep{Molnar}.

\subsubsection{Maxwellian}
A Maxwellian flux distribution is a reasonable simplification for well-moderated neutron sources, and it has the added benefit of being straightforward to integrate in the calculation of Westcott $g$ factors. For a Maxwellian neutron source with temperature $T$, the normalized neutron flux $\phi_{T}(v)$ is given by
\begin{equation}
	\phi_{T}(v)=2\frac{v^{3}}{v_{T}^{4}}e^{-\frac{v^{2}}{v_{T}^{2}}}\label{eq:maxwellian_flux},
\end{equation}
where $v_{T}=\sqrt{2kT/m_{n}}$ and $k$ is the Boltzmann constant. 

When the neutron flux spectrum is known, a Maxwellian distribution with temperature $T$ can be chosen such that it most closely matches the actual neutron spectrum~\citep{Chand25}. This value of $T$ is then employed to look up the Westcott $g$ factor in tables such as those in Ref.~\citep{IAEA} or in Table~\ref{Table:Maxwellian_g-factors} of the Appendix to this paper. In general, however, the flux profiles for most neutron sources do not follow pure Maxwellian distributions. For sources that employ a beam guide, which internally reflects neutrons below a certain wavelength, the flux distribution can be approximated more accurately using the ideal guide spectrum described below.

\subsubsection{Ideal Neutron Guide}
Neutron guides, such as those employed in the beamlines at the BRR and FRM~II, operate according to the principle of total internal reflection. For neutrons with a wavelength $\lambda$, the transmission through a beam guide increases in proportion to $\lambda^2$~\citep{Molnar}. The resulting neutron flux distribution, expressed as a function of $\lambda$, follows that of a Maxwellian distribution multiplied by $\lambda^2$:
\begin{equation}
	\phi_{T}(\lambda)=2\frac{\lambda_{T}^{2}}{\lambda^{3}}e^{-\frac{\lambda_{T}^{2}}{\lambda^{2}}}\label{eq:ideal_guide_flux_lambda},
\end{equation}
where $\lambda_{T}=\sqrt{h^{2}/2m_{n}kT}$. When applying the variable substitution $v = \frac{h}{m_{n}\lambda}$ to put this distribution in terms of the neutron velocity $v$, the ideal-guide flux distribution becomes
\begin{equation}
	\phi_{T}(v)=2\frac{v}{v_{T}^{2}}e^{-\frac{v^{2}}{v_{T}^{2}}}\label{eq:ideal_guide_flux_velocity}.
\end{equation}
The ideal neutron guide spectrum is a useful representation of some neutron beams, although in practice one still must estimate the temperature $T$ used to define the distribution in Eq.~\ref{eq:ideal_guide_flux_velocity}. A table of ideal-guide $g$ factors is included in this paper as Table~\ref{Table:Ideal_guide_g-factors} of the Appendix.

A general and more accurate method for evaluating $g$ factors is to integrate over the measured neutron flux spectrum, instead of using a Maxwellian or ideal-guide approximation to the spectrum. The methodology for evaluating Westcott $g$ factors for arbitrary neutron spectra (excluding epithermal and higher-energy sources) follows directly integrating Eqs.~\ref{eq:gfactor_integral_velocity}, \ref{eq:gfactor_integral_energy}, or \ref{eq:gfactor_integral_wavelength}, and spectra from the PGAA beamlines at the BRR and FRM~II were used as examples for performing these calculations. Particularly for the BRR spectra, the spectra were sufficiently different from a Maxwellian shape that the method of matching a Maxwellian distribution to the actual spectra was insufficient to derive an accurate value for $g$ to apply with these neutron sources.  The BRR and FRM-II neutron spectra are described in further detail below.

\subsubsection{BRR} At the BRR, the neutron beamline from the reactor to the PGAA station can be cooled in a liquid H$_{2}$ cold cell prior to transiting a nickel supermirror guide, which removes the epithermal component from the neutron spectrum at the target position. In the 30-m distance between the cold cell and the target, neutron temperatures increase to an average energy of 12~meV, corresponding to a temperature of 140~K at the target, as measured by time-of-flight techniques~\citep{Belgya12,Belgya14}. The BRR neutron flux distribution $\phi(v)$ from such a measurement~\citep{Belgya14} is shown in Fig.~\ref{fig:Spectra}, along with that of an older spectrum from the BRR operated without the cold source~\citep{Revay04}. Maxwellian neutron flux distributions defined according to Eq.~\ref{eq:maxwellian_flux} with $T=140$~K and $T=293$~K are included in Fig.~\ref{fig:Spectra}(a) for comparison with the measured distributions.  

\subsubsection{FRM~II} 
\begin{figure}[pos=h!]
	\centering{}\includegraphics[width=0.75\textwidth]{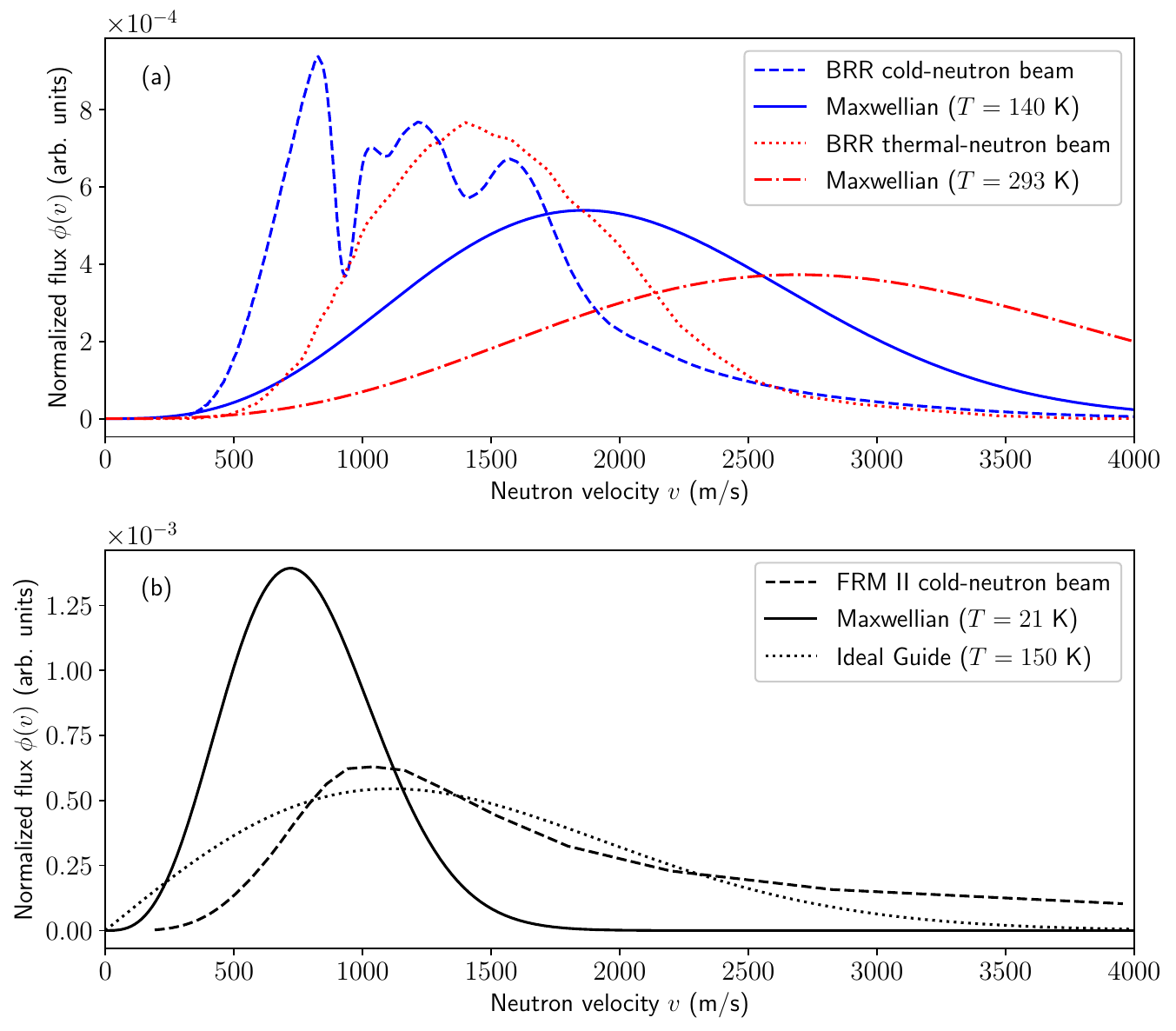}\caption{Normalized neutron flux distributions $\phi(v)$ for guided cold- and thermal-neutron beams from the PGAA beamlines at the Budapest Research Reactor~\citep{Belgya12, Belgya14} (a) and the FRM~II reactor~\citep{Kudejova08} (b), compared with Maxwellian and ideal-guide distributions $\phi_{T}(v)$ defined with various average temperatures $T$. \protect\label{fig:Spectra}}
\end{figure}

The cold-neutron PGAA beamline at FRM~II also employs liquid deuterium cooling and a supermirror guide.  The source spectrum has an average neutron energy of 1.8~meV~\citep{Kudejova08}, corresponding to a temperature of $T=E/k=21$~K.  A Maxwellian neutron flux distribution defined according to Eq.~\ref{eq:maxwellian_flux} with $T=21$~K, and an ideal-guide neutron spectrum defined according to Eq.~\ref{eq:ideal_guide_flux_velocity} with $T=150$~K, are included in Fig.~\ref{fig:Spectra}(b) for comparison with the measured distribution.

\FloatBarrier 
It is evident from Fig.~\ref{fig:Spectra}(a) that Maxwellian distributions are poor representations of the BRR neutron flux distributions, particularly if one attempts to use a thermal ($T=293$~K) Maxwellian distribution to represent the reactor spectrum without the cold source operational. The disagreement between the BRR cold-neutron source spectrum and a Maxwellian distribution with $T=140$~K is also apparent, even though the measured average beam temperature was 140~K.  The flux distribution of the cold-neutron beam at FRM~II PGAA facility more closely resembles a Maxwellian distribution, but the Maxwellian with an temperature of 21~K is noticeably narrower and shifted lower in energy than the measured distribution in Fig.~\ref{fig:Spectra}(b). However, an ideal-guide distribution with a temperature of $T=150$~K is a closer approximation to the FRM~II spectrum. In Sec.~\ref{subsec:arbitrary_spectra_results}, we will show how these differences in neutron spectra can have a marked impact on the calculated Westcott $g$ factors for non-$1/v$ nuclei.

\section{Results\protect\label{sec:Results}}
The methods for calculating Westcott $g$ factors described in Sec.~\ref{sec:Calculations} were compared in several steps to identify differences between the methods.  The irregular isotopes used for these comparisons were the same as the set of isotopes in Table~\ref{Table:Resonance_parameters}, spanning the space from nearly-$1/v$ to highly irregular behavior in their low-energy neutron-capture cross sections.

\subsection{Comparison Between Irregularity Function Method and References}
As described in Sec.~\ref{subsec:Irreg_funcs}, the Westcott $g$ factor calculations in this paper used the lowest-energy resonance (highlighted in bold text in Table~\ref{Table:Resonance_parameters}) for each nucleus to define the irregularity function of Eq.~\ref{eq:lorentzian_single}. The results of these calculations are shown in Table~\ref{Table:References_comparison}, alongside the published reference Westcott $g$ factors from the IAEA Database~\citep{IAEA} and the PGAA Handbook~\citep{Molnar} for Maxwellian neutron flux distributions with various average temperatures $T$, where available. The columns labeled `ENDF' and `Atlas' in Table~\ref{Table:References_comparison} list the results from calculating Westcott $g$ factors with the resonance parameters from the respective source highlighted in Table~\ref{Table:Resonance_parameters}.

\begin{table}[pos=h!]
	\centering{}\caption{Westcott $g$ factors calculated with the irregularity function method, with parameters from ENDF/B-VIII.1~\citep{Brown18,Nobre24} and the \textit{Atlas of Neutron Resonances}~\citep{Mughabghab18} for select non-$1/v$ nuclei, using Maxwellian neutron energy distributions with varying average temperatures $T$.  Included for comparison are the reference Westcott $g$ factor values from the IAEA \textit{Database of Prompt Gamma Rays from Slow Neutron Capture for Elemental Analysis}~(\citep{IAEA}) and the PGAA Handbook~(\citep{Molnar}). \protect\label{Table:References_comparison}}
	\begin{tabular}{|p{1cm}||p{1.5cm}|p{2.6cm}|p{2cm}|p{2cm}|p{2cm}|}
		\hline
		Isotope & $T$ (K) & $g_T$(ENDF~\citep{Brown18,Nobre24}) & $g_T$(Atlas~\citep{Mughabghab18}) & $g_T$(IAEA~\citep{IAEA}) & $g_T$(PGAA~\citep{Molnar}) \\
		\hline
		\hline
		$^{83}$Kr & 100 & 1.003 & 1.003 & 1.006 & 1.006\\
		& 200 & 1.000 & 1.000 & 1.000 & 1.000\\
		& 293 & 0.997 & 0.997 & 0.995 & 0.994\\
		& 400 & 0.995 & 0.995 & 0.988 &  \\
		& 500 & 0.992 & 0.992 & 0.982 &  \\
		& 600 & 0.990 & 0.990 & 0.976 &  \\
		\hline
		$^{115}$In & 100 & 0.983 & 0.983 & 0.984 & 0.983\\
		& 200 & 1.001 & 1.001 & 1.002 & 1.002\\
		& 293 & 1.019 & 1.019 & 1.019 & 1.019\\
		& 400 & 1.041 & 1.041 & 1.041 &  \\
		& 500 & 1.062 & 1.062 & 1.062 &  \\
		& 600 & 1.085 & 1.085 & 1.084 &  \\
		\hline
		$^{149}$Sm & 100 & 0.797 & 0.797 & 0.800 & 0.799\\
		& 200 & 1.227 & 1.230 & 1.239 & 1.245\\
		& 293 & 1.694 & 1.702 & 1.718 & 1.738\\
		& 400 & 2.088 & 2.100 & 2.119 &  \\
		& 500 & 2.290 & 2.305 & 2.325 &  \\
		& 600 & 2.372 & 2.389 & 2.409 &  \\
		\hline
		$^{151}$Eu & 100 & 1.266 & 1.201 & 1.161 & 1.128$^{a}$\\
		& 200 & 1.059 & 1.010 & 1.010 & 0.974$^{a}$\\
		& 293 & 0.919 & 0.865 & 0.900 & 0.843$^{a}$\\
		& 400 & 0.798 & 0.736 & 0.811 &  \\
		& 500 & 0.710 & 0.642 & 0.761 &  \\
		& 600 & 0.639 & 0.567 & 0.743 &  \\
		\hline
		$^{157}$Gd & 100 & 0.883 & 0.881 & 0.887 & 0.881\\
		& 200 & 0.898 & 0.897 & 0.899 & 0.896\\
		& 293 & 0.853 & 0.852 & 0.852 & 0.851\\
		& 400 & 0.782 & 0.780 & 0.779 &  \\
		& 500 & 0.713 & 0.712 & 0.710 &  \\
		& 600 & 0.651 & 0.649 & 0.647 &  \\
		\hline
		$^{176}$Lu & 100 & 0.845 & 0.843 & 0.847 & 0.995$^{b}$\\
		& 200 & 1.167 & 1.168 & 1.176 & 1.000$^{b}$\\
		& 293 & 1.714 & 1.699 & 1.752 & 1.003$^{b}$\\
		& 400 & 2.463 & 2.396 & 2.545 &  \\
		& 500 & 3.080 & 2.953 & 3.205 &  \\
		& 600 & 3.545 & 3.362 & 3.704 &  \\
		\hline
	\end{tabular}
	\\$^{a}$Values could only be reproduced using $E_r = +0.00361$ eV; see note in Table \ref{Table:Resonance_parameters}.\\
	$^{b}$Values could only be reproduced using $E_r = 4.36$ eV and $\Gamma = 68.4$ meV; see note in Table \ref{Table:Resonance_parameters}.
\end{table}

It is evident in Table~\ref{Table:References_comparison} that the irregularity function method reliably reproduces the published data in Ref.~\citep{IAEA} to within 1\% in most cases, and that the Westcott $g$ factors are not especially sensitive to the source (\mbox{ENDF/B-VIII.1} or \textit{Atlas of Neutron Resonances}) of the resonance parameters used. The methodology described in the IAEA Database involves using the ENDF utility code \texttt{INTER}~\citep{INTER} to generate Westcott $g$ factors by direct integration of the neutron cross sections. Despite using different methods, the two references arrive at similar values of $g$ for the same Maxwellian temperatures, except for the highly irregular isotope $^{176}$Lu, where differences grow to 4.5\% at 600~K between the calculated $g$ factors and those in the IAEA Database.

The irregularity function method described in Sec.~\ref{subsec:Irreg_funcs} was used to create the table of Westcott $g$ factors in the PGAA Handbook~\citep{Molnar}, where the resonance parameters were taken from ENDF/B-VI.8~\citep{Dunford92} and are slightly different than those in ENDF/B-VIII.1. By using the resonance parameters in the PGAA Handbook~\cite{Molnar} listed in Table~\ref{Table:Resonance_parameters}, we were able to reproduce the $g$ factors listed in that reference using the irregularity function method, with two exceptions. The $g$ factors in the PGAA Handbook differ from the calculated values for $^{151}$Eu and $^{176}$Lu: for $^{151}$Eu, these differences were attributed to a missing sign in the resonance parameters in~\citep{Molnar}, and for $^{176}$Lu it was likely due to the use of a higher-energy resonance to calculate the values in the PGAA Handbook.

\subsection{Comparison Between Irregularity Function and Cross-Section Integration Methods}

The irregularity function method described in Sec.~\ref{subsec:Irreg_funcs} was then compared to the cross-section integration method outlined in Sec.~\ref{subsec:CS_integration}, by calculating Westcott $g$ factors for the same set of comparator isotopes as in Table~\ref{Table:References_comparison}.  The results of these calculations, which were performed using the \texttt{WestcottFactors} toolkit developed as part of this work, are shown in Table~\ref{Table:Irregularity_comparison}.  Here, resonance parameters and cross sections were taken from ENDF/B-VIII.1~\citep{Brown18,Nobre24} for both sets of $g$-factor calculations. For comparison, the values of $g_T$ from the IAEA \textit{Database of Prompt Gamma Rays from Slow Neutron Capture for Elemental Analysis}~(\citep{IAEA}) and the PGAA Handbook~(\citep{Molnar}) for Maxwellian neutron spectra with the same temperatures $T$ are included. 

\begin{table}[pos=h!]
	\centering{}\caption{Westcott $g$ factors calculated for select non-$1/v$ nuclei using Maxwellian neutron flux distributions with average temperatures $T$ between 100-600~K.  For the irregularity function method, resonance parameters are from ENDF/B-VIII.1, and for the cross-section integration method, cross sections were taken from GNDS-formatted data in ENDF/B-VIII.1.  Percentage differences between the $g$ values obtained using the two methods are included to facilitate comparison. \protect\label{Table:Irregularity_comparison}}
	\begin{tabular}{|p{1cm}||p{1cm}|p{3.5cm}|p{4cm}|p{2cm}|}
		\hline
		Isotope & $T$ (K) & $g_T$(Irregularity Function) & $g_T$(Cross Section Integration) & \% Difference \\
		\hline
		\hline
		$^{83}$Kr & 100 & 1.003 & 1.003 & 0.0\\
		& 200 & 1.000 & 1.000 & 0.0\\
		& 293 & 0.997 & 0.998 & 0.0\\
		& 400 & 0.995 & 0.995 & 0.0\\
		& 500 & 0.992 & 0.993 & 0.1\\
		& 600 & 0.990 & 0.990 & 0.1\\
		\hline
		$^{115}$In & 100 & 0.983 & 0.983 & 0.0\\
		& 200 & 1.001 & 1.001 & 0.0\\
		& 293 & 1.019 & 1.019 & -0.0\\
		& 400 & 1.041 & 1.040 & -0.0\\
		& 500 & 1.062 & 1.062 & -0.1\\
		& 600 & 1.085 & 1.084 & -0.1\\
		\hline
		$^{149}$Sm & 100 & 0.797 & 0.801 & 0.5\\
		& 200 & 1.227 & 1.222 & -0.4\\
		& 293 & 1.694 & 1.679 & -0.9\\
		& 400 & 2.088 & 2.064 & -1.1\\
		& 500 & 2.290 & 2.262 & -1.2\\
		& 600 & 2.372 & 2.342 & -1.3\\
		\hline
		$^{151}$Eu & 100 & 1.266 & 1.216 & -4.1\\
		& 200 & 1.059 & 1.051 & -0.7\\
		& 293 & 0.919 & 0.946 & 2.8\\
		& 400 & 0.798 & 0.863 & 7.6\\
		& 500 & 0.710 & 0.818 & 13.2\\
		& 600 & 0.639 & 0.802 & 20.3\\
		\hline
		$^{157}$Gd & 100 & 0.883 & 0.884 & 0.2\\
		& 200 & 0.898 & 0.898 & 0.0\\
		& 293 & 0.853 & 0.853 & -0.0\\
		& 400 & 0.782 & 0.781 & -0.1\\
		& 500 & 0.713 & 0.712 & -0.1\\
		& 600 & 0.651 & 0.650 & -0.2\\
		\hline
		$^{176}$Lu & 100 & 0.845 & 0.847 & 0.2\\
		& 200 & 1.167 & 1.164 & -0.2\\
		& 293 & 1.714 & 1.704 & -0.6\\
		& 400 & 2.463 & 2.442 & -0.9\\
		& 500 & 3.080 & 3.050 & -1.0\\
		& 600 & 3.545 & 3.508 & -1.0\\
		\hline
	\end{tabular}
\end{table}

While the Westcott $g$ factors in Table~\ref{Table:Irregularity_comparison} show little difference between the irregularity function and cross-section integration methods for some isotopes, such as $^{83}$Kr and $^{115}$In, these isotopes also have the closest to $1/v$ behavior of the irregular nuclei used in this comparison. This is evident from the fact that their Westcott $g$ factors are within 1\% percent of unity for Maxwellian temperatures between 100-600~K. 

For other, more irregular nuclei such as $^{149}$Sm, $^{151}$Eu, $^{157}$Gd, and $^{176}$Lu, which have Westcott $g$ factors that differ substantially from unity, the method used to calculate $g$ factors can have a larger effect. For instance, at Maxwellian temperatures between 200-600~K, the irregularity function method produces Westcott $g$ factors for $^{151}$Eu that are up to 20\% lower than those obtained by integrating over the full neutron-capture cross section. These results suggest that the assumption employed in the irregularity function method, where the lowest-energy resonance dominates the capture cross section at low neutron energies, is an oversimplification that is only valid for $1/v$ or near-$1/v$ nuclei.  The more accurate method of integrating over the cross sections in the ENDF/B-VIII.1 libraries, which is employed in \texttt{DeCE} and the \texttt{WestcottFactors} tool described in this paper, should be used for nuclei with multiple low-energy resonances in their neutron-capture cross sections.

\subsection{Extension to Arbitrary Non-Maxwellian Neutron Spectra\protect\label{subsec:arbitrary_spectra_results}}
The \texttt{WestcottFactors} toolkit natively incorporates functionality to ingest user-defined, CSV-formatted neutron energy spectra to calculate Westcott $g$ factors, and a similar capability resides in the \texttt{DeCE} development branch fork on GitHub ~\citep{DeCE_fork_develop-GitHub}. The public GitHub and PyPI repositories for the \texttt{WestcottFactors} code ~\citep{WestcottFactors-GitHub, WestcottFactors-PyPI} include the experimental BRR and FRM~II spectra described in Sec.~\ref{subsec:Neutron_dist}, which have been used to calculate Westcott $g$ factors with cross sections from ENDF/B-VIII.1. The results are shown in Table~\ref{Table:Reactor_comparison}. 

\begin{table}[pos=h!]
	\renewcommand{\arraystretch}{1.2}
	\centering{}\caption{Comparison of Westcott $g$ factors for select non-$1/v$ nuclei using Maxwellian neutron flux distributions, BRR cold- and thermal-neutron spectra, and the FRM-II cold-neutron spectrum. \protect\label{Table:Reactor_comparison}}
	\begin{tabular}{|p{1cm}||p{1.5cm}|p{2cm}||p{1.5cm}|p{2.5cm}||p{1.5cm}|p{2.5cm}|}
		\hline
		Isotope & $g_{T=140~\text{K}}$ & $g$, BRR cold & $g_{T=293~\text{K}}$ & $g$, BRR thermal & $g_{T=21~\text{K}}$ & $g$, FRM~II cold \\
		\hline
		\hline
		$^{83}$Kr & 1.002 & 1.004 & 0.998 & 1.001 & 1.005 & 1.003\\
		\hline
		$^{115}$In & 0.991 & 0.977 & 1.019 & 1.054 & 0.970 & 0.982\\
		\hline
		$^{149}$Sm & 0.942 & 0.721 & 1.679 & 0.767 & 0.630 & 0.842\\
		\hline
		$^{151}$Eu & 1.142 & 1.296 & 0.946 & 1.225 & 1.405 & 1.245\\
		\hline
		$^{157}$Gd & 0.902 & 0.846 & 0.853 & 0.880 & 0.792 & 0.852\\
		\hline
		$^{176}$Lu & 0.944 & 0.789 & 1.704 & 0.829 & 0.719 & 0.863\\
		\hline
	\end{tabular}
\end{table}

A comparison between the Westcott $g$ factors for the BRR and FRM~II neutron sources and Maxwellian spectra with average temperatures that might be used to describe these sources reveals differences of 20\% or more for $^{149}$Sm, $^{151}$Eu, $^{157}$Gd, and $^{176}$Lu, suggesting that $g$ factors for non-$1/v$ nuclei depend strongly on the specific neutron spectrum used in activation measurements. Because Maxwellian spectra are theoretical and do not generally resemble the actual neutron spectra used in most thermal- and cold-neutron activation experiments, values for $g$ used to adjust thermal-neutron cross sections should be calculated with the measured neutron flux whenever the spectrum is available. The \texttt{WestcottFactors} package~\citep{WestcottFactors-GitHub, WestcottFactors-PyPI} and the modified \texttt{DeCE} fork~\citep{DeCE_fork_develop-GitHub} both provide users with a straightforward way to perform these calculations with any neutron energy spectrum.

\subsection{Comparison Between \texttt{DeCE} and \texttt{WestcottFactors}}

Next, calculations of Westcott $g$ factors using ENDF/B-VIII.1 cross sections, as described in Sec.~\ref{subsec:CS_integration}, were carried out using both \texttt{DeCE} and the \texttt{WestcottFactors} tool. The results of running the Westcott $g$ factor calculator in the \texttt{DeCE} development fork~\citep{DeCE_fork_develop-GitHub} are listed alongside the corresponding results from running the \texttt{WestcottFactors} tool in Table~\ref{Table:DeCE_comparison} for Maxwellian neutron flux distributions with various average temperatures $T$. Additionally, calculations using the \texttt{DeCE} development fork were compared against similar calculations for the BRR and FRM~II spectra using the \texttt{WestcottFactors} tool. The non-$1/v$ nuclei used in the comparison are the same as those listed in Table~\ref{Table:Resonance_parameters}.

\begin{table}[pos=h!]
	\renewcommand{\arraystretch}{0.9}
	\centering{}\caption{Westcott $g$ factors calculated for select non-$1/v$ nuclei using Maxwellian neutron flux distributions with various average temperatures $T$, as well as the reactor spectra described in Sec.~\ref{subsec:Neutron_dist}.  Cross sections were taken from ENDF/B-VIII.1, in the GNDS format for use with the \texttt{WestcottFactors} tool described in the text, and in the ENDF-6 format for use with \texttt{DeCE}.  Percentage differences between the $g$ values obtained using the two methods are included to facilitate comparison. \protect\label{Table:DeCE_comparison}}
	\begin{tabular}{|p{1cm}||p{2.5cm}|p{3cm}|p{2cm}|p{2cm}|}
		\hline
		Isotope & $T$ (K) or Spectrum & $g$(\texttt{WestcottFactors}) & $g$(\texttt{DeCE}) & \% Difference \\
		\hline
		\hline
		$^{83}$Kr & 100 & 1.003 & 1.004 & -0.1\\
		& 200 & 1.000 & 1.001 & -0.1\\
		& 293 & 0.998 & 0.999 & -0.1\\
		& 400 & 0.995 & 0.997 & -0.1\\
		& 500 & 0.993 & 0.994 & -0.1\\
		& 600 & 0.990 & 0.992 & -0.1\\
		& BRR cold & 1.004 & 1.005 & -0.1\\
		& BRR thermal & 1.001 & 1.002 & -0.2\\
		& FRM II cold & 1.003 & 1.004 & -0.1\\
		\hline
		$^{115}$In & 100 & 0.983 & 0.985 & -0.2\\
		& 200 & 1.001 & 1.003 & -0.1\\
		& 293 & 1.019 & 1.021 & -0.1\\
		& 400 & 1.040 & 1.042 & -0.1\\
		& 500 & 1.062 & 1.063 & -0.1\\
		& 600 & 1.084 & 1.085 & -0.1\\
		& BRR cold & 0.977 & 0.979 & -0.2\\
		& BRR thermal & 1.054 & 1.060 & -0.5\\
		& FRM II cold & 0.982 & 0.984 & -0.2\\
		\hline
		$^{149}$Sm & 100 & 0.801 & 0.803 & -0.3\\
		& 200 & 1.222 & 1.226 & -0.3\\
		& 293 & 1.679 & 1.684 & -0.3\\
		& 400 & 2.064 & 2.069 & -0.3\\
		& 500 & 2.262 & 2.267 & -0.2\\
		& 600 & 2.342 & 2.348 & -0.2\\
		& BRR cold & 0.721 & 0.723 & -0.3\\
		& BRR thermal & 0.767 & 0.770 & -0.3\\
		& FRM II cold & 0.842 & 0.847 & -0.6\\
		\hline
		$^{151}$Eu & 100 & 1.216 & 1.218 & -0.1\\
		& 200 & 1.051 & 1.053 & -0.2\\
		& 293 & 0.946 & 0.948 & -0.2\\
		& 400 & 0.863 & 0.865 & -0.2\\
		& 500 & 0.818 & 0.820 & -0.2\\
		& 600 & 0.802 & 0.804 & -0.2\\
		& BRR cold & 1.296 & 1.298 & -0.1\\
		& BRR thermal & 1.225 & 1.227 & -0.2\\
		& FRM II cold & 1.245 & 1.246 & -0.1\\
		\hline
		$^{157}$Gd & 100 & 0.884 & 0.886 & -0.2\\
		& 200 & 0.898 & 0.900 & -0.2\\
		& 293 & 0.853 & 0.854 & -0.2\\
		& 400 & 0.781 & 0.782 & -0.2\\
		& 500 & 0.712 & 0.714 & -0.2\\
		& 600 & 0.650 & 0.651 & -0.2\\
		& BRR cold & 0.846 & 0.848 & -0.2\\
		& BRR thermal & 0.880 & 0.883 & -0.3\\
		& FRM II cold & 0.852 & 0.854 & -0.2\\
		\hline
		$^{176}$Lu & 100 & 0.847 & 0.849 & -0.2\\
		& 200 & 1.164 & 1.168 & -0.3\\
		& 293 & 1.704 & 1.709 & -0.3\\
		& 400 & 2.442 & 2.448 & -0.3\\
		& 500 & 3.050 & 3.058 & -0.2\\
		& 600 & 3.508 & 3.517 & -0.2\\
		& BRR cold & 0.789 & 0.791 & -0.3\\
		& BRR thermal & 0.829 & 0.832 & -0.3\\
		& FRM II cold & 0.863 & 0.867 & -0.5\\
		\hline
	\end{tabular}
\end{table}

It is clear from Table~\ref{Table:DeCE_comparison} that the modified \texttt{DeCE} code and the \texttt{WestcottFactors} tool described in this paper, which both employ the cross-section integration methodology outlined in Sec.~\ref{subsec:CS_integration}, produce nearly identical Westcott $g$ factors regardless of the neutron spectra used. Because cross-section data from ENDF/B-VIII.1 were used in both the \texttt{DeCE} and \texttt{WestcottFactors} calculations, the small variations of less than 1\% in the $g$ factors result from slight differences in the interpolation and numerical integration methods used in the two codes.

\subsection{Tabulated Westcott $g$ Factors for Irregular Isotopes With Maxwellian and Ideal Guide Distributions}

The close agreement between the Westcott $g$ factors calculated using the modified \texttt{DeCE} code and the\\ \texttt{WestcottFactors} tool provides confidence in the methods used.  \texttt{WestcottFactors} was then employed to produce a table of Maxwellian-spectrum $g_T$ values similar to that in the IAEA database~\citep{IAEA}, which is included in the Appendix as Table \ref{Table:Maxwellian_g-factors}. The set of mostly-irregular isotopes includes those from the IAEA database, as well as those from Refs.~\citep{Molnar, vanSluijs15, Holden99, Pritychenko25}. The intent of publishing this table is to provide PGAA and NAA practitioners with a reference consisting of Westcott $g$ factors for a broad range of Maxwellian-spectrum temperatures, obtained using the most up-to-date neutron-capture cross sections from ENDF/B-VIII.1. Table~\ref{Table:Maxwellian_g-factors} includes several isotopes that display nearly-$1/v$ behavior, evident from $g_T\approx 1$ across the range of temperatures. Because these isotopes appear in other references cited in this paper, they were included here for completeness. It is notable that for several isotopes, including $^{132}${Ba}, $^{138}${Ce}, $^{163}${Dy}, $^{175}${Lu}, $^{180}${Ta}, and $^{186}${Os}, the Westcott $g$-factor values in Table~\ref{Table:Maxwellian_g-factors} differ from the values in the IAEA Database~\citep{IAEA} by greater than 1\%.  Due to the relevance of the ideal-guide flux distribution for approximating guided neutron beams, a table of $g_T$ values for irregular nuclei calculated using the ideal-guide spectrum is also included in the Appendix as Table~\ref{Table:Ideal_guide_g-factors}.

It is worth noting that the Westcott $g$ factors in Tables~\ref{Table:Maxwellian_g-factors} and \ref{Table:Ideal_guide_g-factors} should be treated with caution, particularly for highly-irregular nuclei, for the reasons outlined in Sec.~\ref{subsec:arbitrary_spectra_results}.  If an experimental neutron-energy spectrum is available, it should be used  with the \texttt{WestcottFactors} or \texttt{DeCE} tools described in this work instead of a Maxwellian or ideal guide approximation, to ensure the Westcott $g$ factors are accurate for the particular application.

\FloatBarrier  
\section{Summary\protect\label{sec:Summary}}

Computational methods for evaluating Westcott $g$ factors were developed into the open-source \texttt{WestcottFactors} \citep{WestcottFactors-GitHub, WestcottFactors-PyPI} software package, which allows users to specify an arbitrary neutron spectrum and calculate values of $g$ for use in NAA and PGAA experiments. For users familiar with using \texttt{DeCE} to interact with ENDF data, modifications to that code~\citep{DeCE_fork_develop-GitHub} provide the same functionality for performing these calculations as the \texttt{WestcottFactors} package. Tests of these two software tools demonstrate consistency in their results, so users can choose to use the one that best suits their needs for calculating Westcott $g$ factors.

The results of testing the \texttt{WestcottFactors} package on a select number of irregular non-$1/v$ nuclei show that the Westcott $g$ factors are sensitive to the choice of neutron spectrum, and certain low-energy neutron-capture cross section resonances. In the case of highly-irregular nuclei with multiple resonances below 5~eV, employing irregularity functions to approximate the calculation of Westcott $g$ factors may result in significant differences when compared to the more accurate method of integrating over the full neutron-capture cross sections. 

Most importantly, in this work we showed that Westcott $g$ factors calculated with Maxwellian neutron flux distributions can result in errors when approximating the actual $g$-factor values obtained by integrating over the measured neutron spectra; this was demonstrated using experimentally-measured spectra from guided thermal- and cold-neutron beamlines at BRR and FRM~II. This comparison demonstrates the utility of the open-source \texttt{WestcottFactors} software package, and the modifications to the \texttt{DeCE} code, which have been made available as part of this work. For PGAA and NAA practitioners, these tools provide a capability for confirming tabulated $g$-factor values. It is recommended that even if a tabulated $g$ factor is used, whenever possible it should be verified with either of these software packages using the measured neutron spectrum. As it was shown in this work, even relatively minor differences in the neutron spectrum can produce significant changes in the calculated $g$ factors for certain irregular nuclei, so the tabulated values that assume Maxwellian or ideal-guide neutron flux distributions are inherently less accurate than the $g$ factors calculated using the actual spectrum.

\FloatBarrier 
\section*{Acknowledgments}

This work was supported at the Lawrence Berkeley National Laboratory under Contract No. DE-AC02-05CH11231 and at the Los Alamos National Laboratory under Contract No. 89233218CNA000001 for the U.S. Nuclear Data Program.  Work at the University of California, Berkeley, was supported by the U.S. Department of Energy National Nuclear Security Administration through the Nuclear Science and Security Consortium under Award Number DE-NA0003996. The authors would like to express their gratitude to Laszlo Szentmiklósi, Tamas Belgya, and Zsolt Révay for providing neutron flux data from the Budapest Research Reactor and FRM~II.


\bibliographystyle{elsarticle-num} 

\bibliography{westcott_references}

@book{IAEA,
	address = {Vienna, Austria},
	editor = {Choi, H. D. and others},
	publisher = {International Atomic Energy Agency},
	title = {Database of Prompt Gamma Rays from Slow Neutron Capture for Elemental Analysis},
	year = {2007}}

@book{Mughabghab18,
	author = {Mughabghab, S. F.},
	edition = {6th},
	publisher = {Elsevier Science},
	title = {Resonance Properties and Thermal Cross Sections $Z$=1-102},
	year = {2018}}

@book{Molnar,
	address = {Dordrecht, the Netherlands},
	editor = {Moln{\'a}r, G{\'a}bor L.},
	publisher = {Kluwer Academic},
	title = {Handbook of Prompt Gamma Activation Analysis},
	year = {2004}}

@article{Matters16,
	author = {Matters, D. A. and Lerch, A. G. and Hurst, A. M. and Szentmikl{\'o}si, L. and Carroll, J. J. and R{\'e}vay, Zs. and McClory, J. W. and McHale, S. R. and Firestone, R. B. and Sleaford, B. W. and Krti{\v c}ka, M. and Belgya, T.},
	journal = {Physical Review C},
	pages = {054319},
	title = {Investigation of $^{186}${Re} via radiative thermal-neutron capture on $^{185}${Re}},
	volume = {93},
	year = {2016}}

@article{Hurst19,
	author = {Hurst, A. M. and Sweet, A. and Goldblum, B. L. and Firestone, R. B. and Basunia, M. S. and Bernstein, L. A. and R\'evay, Zs. and Szentmikl\'osi, L. and Belgya, T. and Escher, J. E. and Hars\'anyi, I. and Krti\ifmmode \check{c}\else \v{c}\fi{}ka, M. and Sleaford, B. W. and Vujic, J.},
	journal = {Phys. Rev. C},
	pages = {024310},
	title = {Radiative-capture cross sections for the $^{139}\mathrm{La}(n,\ensuremath{\gamma})$ reaction using thermal neutrons and structural properties of $^{140}\mathrm{La}$},
	volume = {99},
	year = {2019}}

@article{Choi14,
	author = {Choi, H. D. and Firestone, R. B. and Basunia, M. S. and Hurst, A. and Sleaford, B. and Summers, N. and Escher, J. E. and R{\'e}vay, Zs. and Szentmikl{\'o}si, L. and Belgya, T. and Krti{\v c}ka, M.},
	journal = {Nuclear Science and Engineering},
	pages = {219--232},
	title = {Radiative capture cross sections of $^{155,157}${Gd} for thermal neutrons},
	volume = {177},
	year = {2014}}

@techreport{Firestone21,
	address = {Vienna, Austria},
	author = {Firestone, R. B.},
	institution = {International Atomic Energy Agency},
	month = {October},
	number = {INDC(USA)-109},
	publisher = {IAEA},
	title = {Renormalization of Pile Oscillator Thermal Neutron Capture Cross Section Data},
	year = {2021}}

@article{Trkov20,
	author = {Trkov, A.. and others},
	journal = {Nuclear Data Sheets},
	pages = {1--108},
	title = {{IRDFF-II}: A New Neutron Metrology Library},
	volume = {163},
	year = {2020}}

@url{EGAF,
	author = {Firestone, R. B.},
	title = {Database of prompt gamma rays from slow neutron capture for elemental analysis, {(International Atomic Energy Agency, Vienna, 2006)}},
	url = {https://www-nds.iaea.org/pgaa/egaf.html}}

@article{Hurst15,
	author = {Hurst, A. M. and Summers, N. C. and Szentmikl{\'o}si, L. and Firestone, R. B. and Basunia, M. S. and Escher, J. E. and Sleaford, B. W.},
	journal = {Nuclear Instruments and Methods in Physics Research B},
	pages = {38--44},
	title = {Determination of the effective sample thickness via radiative capture},
	volume = {362},
	year = {2015}}

@article{Hurst23,
	author = {Hurst, A. M. and Firestone, R. B. and Chimanski, E. V.},
	journal = {Nuclear Instruments and Methods in Physics Research A},
	pages = {168715},
	title = {py{EGAF}: An open-source {Python} library for the evaluated gamma-ray activation file},
	volume = {1067},
	year = {2023}}

@url{pyEGAF-pypi,
	author = {Hurst, A. M.},
	title = {{pyEGAF} ({PyPI} package)},
	url = {https://pypi.org/project/pyEGAF/},
	urldate = {2023}}

@article{Westcott55,
	author = {Westcott, C. H.},
	journal = {Journal of Nuclear Energy},
	pages = {59 -- 76},
	title = {The specification of neutron flux and nuclear cross-sections in reactor calculations},
	volume = {2},
	year = {1955}}

@techreport{Westcott60,
	address = {Chalk River, Ontario},
	author = {Westcott, C. H.},
	institution = {Atomic Energy of Canada Limited},
	number = {CRRP-960},
	title = {Effective cross section values for well-moderated thermal reactor spectra},
	year = {1960}}

@article{Brown18,
	author = {Brown, D. A. and others},
	journal = {Nuclear Data Sheets},
	pages = {1--142},
	title = {{ENDF/B-VIII.0}: The 8th Major Release of the Nuclear Reaction Data Library with {CIELO}-project Cross Sections, New Standards and Thermal Scattering Data},
	volume = {148},
	year = {2018}}

@book{Byrne,
	address = {Bristol and Philadelphia},
	editor = {Byrne, J.},
	publisher = {Institute of Physics Publishing},
	title = {Neutrons, Nuclei, and Matter},
	year = {1994}}

@article{Nobre24,
	author = {Nobre, G. and Brown, D. and Arcilla, R. and Coles, R. and Shu, B.},
	journal = {European Physics Journal Web of Conferences},
	pages = {04004},
	title = {Progress towards the {ENDF/B-VIII.1} release},
	volume = {294},
	year = {2024}}

@article{Kawano19,
	author = {Kawano, Toshihiko},
	journal = {Journal of Nuclear Science and Technology},
	number = {11},
	pages = {1029 -- 1035},
	title = {{DeCE}: the {ENDF-6} data interface and nuclear data evaluation assist code},
	volume = {56},
	year = {2019}}

@url{DeCE-GitHub,
	author = {Kawano, T.},
	title = {{DeCE} ({GitHub} repository)},
	url = {https://github.com/toshihikokawano/DeCE},
	urldate = {2025}}

@url{DeCE_fork_develop-GitHub,
	author = {Matters, D. and Kawano, T.},
	title = {Fork to {DeCE} Develop Branch ({GitHub} repository)},
	url = {https://github.com/DMatters/DeCE_g-Factor/tree/develop},
	urldate = {2025}}

@techreport{Brown20,
	author = {Brown, D. and Beck, B. and Mattoon, C. and Bailey, T. and Thompson, I. and Conlin, J. L. and Haeck, W. and White, M. and Paris, M. and Fleming, M. and others},
	institution = {Organisation for Economic Cooperation and Development},
	number = {Technical Report NEA--7519},
	title = {Specifications for the {G}eneralised {N}uclear {D}atabase {S}tructure ({GNDS})-Version 1.9},
	year = {2020}}

@url{WestcottFactors-GitHub,
	author = {Matters, D. A. and Hurst, A. M.},
	title = {Westcott Factors ({GitHub} repository)},
	url = {https://github.com/DMatters/WestcottFactors},
	urldate = {2025}}

@url{WestcottFactors-PyPI,
	author = {Matters, D. A. and Hurst, A. M.},
	title = {{WestcottFactors} ({PyPI} package)},
	url = {https://pypi.org/project/westcott/},
	urldate = {2026}}

@techreport{Brown23,
	author = {Brown, D. A. and others},
	editor = {Brown, D. A.},
	institution = {Cross Sections Evaluations Working Group},
	month = {September},
	number = {BNL-224854-2023-INR},
	title = {{ENDF-6} Formats Manual},
	year = {2023}}

@inproceedings{McKinney10,
	author = {McKinney, Wes},
	booktitle = {Proceedings of the 9th {P}ython in Science Conference},
	editor = {van der Walt, St\'efan and Millman,Jarrod},
	pages = {56 - 61},
	title = {Data Structures for Statistical Computing in {P}ython},
	year = {2010}}

@article{Mattoon23,
	author = {Mattoon, C. and Beck, B. and Godfree, G.},
	journal = {EPJ Web of Conferences},
	pages = {14010},
	title = {Managing and Processing Nuclear Data Libraries with {FUDGE}},
	volume = {284},
	year = {2023}}

@url{FUDGE-GitHub,
	author = {Mattoon, C.},
	title = {For Updating Data and Generating Evaluations ({FUDGE}): {LLNL} code for managing nuclear data ({GitHub} repository)},
	url = {https://github.com/LLNL/fudge},
	urldate = {2025}}

@article{Virtanen20,
	author = {Virtanen, Pauli and {SciPy 1.0 Contributors}},
	journal = {Nature Methods},
	pages = {261--272},
	title = {{{SciPy} 1.0: Fundamental Algorithms for Scientific Computing in Python}},
	volume = {17},
	year = {2020}}

@incollection{Nichols08,
	address = {Vienna, Austria},
	author = {Nichols, A. L. and Aldama, D. L. and Verpelli, M.},
	booktitle = {Handbook of Nuclear Data for Safeguards},
	institution = {International Atomic Energy Agency},
	month = {August},
	number = {INDC(NDS)-0534},
	publisher = {IAEA},
	title = {Database Extensions, {A}ugust 2008},
	year = {2008}}

@article{vanSluijs15,
	author = {{van Sluijs}, R. and Stopic, A. and Jacimovic, R.},
	journal = {Journal of Radioanalytical and Nuclear Chemistry},
	pages = {579--587},
	title = {Evaluation of {Westcott} $g(T_n)$-factors used in $k_0$-{NAA} for "non-$1/v$" $(n,\gamma)$ reactions},
	volume = {306},
	year = {2015}}

@article{Holden99,
	author = {Holden, N. E.},
	journal = {Pure and Applied Chemistry},
	number = {12},
	pages = {2309--2315},
	title = {Temperature dependence of the {W}estcott $g$-factor for neutron reactions in activation analysis ({IUPAC} technical report)},
	volume = {71},
	year = {1999}}

@article{Pritychenko25,
	author = {Pritychenko, B.},
	journal = {Atomic Data and Nuclear Data Tables},
	pages = {101708},
	title = {Tables of neutron thermal cross sections, {W}estcott factors, resonance integrals, {M}axwellian averaged cross sections, astrophysical reaction rates, and $r$-process abundances calculated from the {ENDF/B-VIII.1}, {JEFF-3.3}, {JENDL-5.0}, {BROND-3.1}, and {CENDL-3.2} evaluated data libraries},
	volume = {163},
	year = {2025}}

@article{Chand25,
	author = {Chand, M. and Bagchi, S. and Khan, B. H.},
	journal = {Applied Radiation and Isotopes},
	pages = {111666},
	title = {Determination of {W}estcott $g$-factors for the assay of non-$1/v$ nuclides using $k_0$-{NAA}},
	volume = {217},
	year = {2025}}

@article{Belgya12,
	author = {Belgya, T.},
	journal = {Physics Procedia},
	pages = {99--109},
	title = {Prompt gamma activation analysis at the {Budapest Research Reactor}},
	volume = {31},
	year = {2012}}

@article{Belgya14,
	author = {Belgya, T. and Kis, Z. and Szentmikl{\'o}si, L.},
	journal = {Nuclear Data Sheets},
	pages = {419--421},
	title = {Neutron flux characterization of the cold beam {PGAA-NIPS} facility at the {B}udapest research reactor},
	volume = {119},
	year = {2014}}

@article{Revay04,
	author = {R{\'e}vay, Zs. and Belgya, T. and Kasztovszky, Zs. and Weil, J. L. and Moln{\'a}r, G. L.},
	journal = {Nuclear Instruments and Methods in Physics Research B},
	pages = {385--388},
	title = {Cold neutron {PGAA} facility at {Budapest}},
	volume = {213},
	year = {2004}}

@article{Kudejova08,
	author = {Kudejova, P. and Meierhofer, G. and Zeitelhack, K. and Jolie, J. and Schulze, R. and T{\"u}rler, A. and Materna, T.},
	journal = {Journal of Radioanalytical and Nuclear Chemistry},
	number = {3},
	pages = {691--695},
	title = {The new {PGAA} and {PGAI} facility at the research reactor {FRM II} in {G}arching near {M}unich},
	volume = {278},
	year = {2008}}

@url{INTER,
	author = {Dunford, C. L. and {Organisation for Economic Co-Operation and Development, Nuclear Energy Agency - OECD/NEA}},
	title = {INTER: {ENDF/B} Thermal Cross-Sections, Resonance Integrals, G-Factors Calculation},
	url = {https://inis.iaea.org/records/fqpqg-d3r21}}

@inproceedings{Dunford92,
	author = {Dunford, C. L.},
	address = {Berlin, Heidelberg},
	booktitle = {Nuclear Data for Science and Technology},
	editor = {Qaim, Syed M.},
	isbn = {978-3-642-58113-7},
	pages = {788--792},
	publisher = {Springer Berlin Heidelberg},
	title = {Evaluated Nuclear Data File, {ENDF/B-VI}},
	year = {1992}}

\clearpage
\section*{Appendix: Maxwellian and Ideal-Guide Spectrum Westcott $g$ Factors for Irregular Nuclei\protect\label{Appendix_Maxwellian_gfactors}}

\setcounter{table}{0}
\renewcommand{\thetable}{A\arabic{table}}
	
\begin{table}[pos=ht!]
	\centering{}\caption{Maxwellian-distribution Westcott $g_T$ factors at various temperatures $T$ for non-$1/v$ nuclei listed in References~\citep{IAEA, Molnar, vanSluijs15, Holden99, Pritychenko25}, calculated using the cross-section integration method in the \texttt{WestcottFactors} toolkit~\citep{WestcottFactors-GitHub, WestcottFactors-PyPI}, with cross sections taken from ENDF/B-VIII.1~\citep{Brown18,Nobre24}. \protect\label{Table:Maxwellian_g-factors}}
	\renewcommand{\arraystretch}{1.5}
\begin{tabular}{|c||c|c|c|c|c|c|c|c|c|c|c|c|c|}
		\hline
		$T~(K)$ & $^{30}$Si & $^{36}$S & $^{36}$Ar & $^{38}$Ar & $^{83}$Kr & $^{87}$Sr & $^{93}$Nb & $^{103}$Rh & $^{105}$Pd & $^{107}$Ag & $^{109}$Ag & $^{111}$Cd & $^{113}$Cd\\
		\hline
		\hline
		20 & 1.000 & 0.775 & 1.141 & 1.247 & 1.005 & 0.990 & 1.000 & 0.964 & 1.002 & 1.003 & 0.992 & 1.001 & 0.782\\
		\hline
		40 & 1.000 & 0.829 & 1.120 & 1.266 & 1.004 & 0.991 & 1.000 & 0.968 & 1.002 & 1.003 & 0.993 & 1.001 & 0.804\\
		\hline
		60 & 1.000 & 0.863 & 1.090 & 1.222 & 1.004 & 0.992 & 1.000 & 0.972 & 1.001 & 1.002 & 0.994 & 1.001 & 0.828\\
		\hline
		80 & 1.000 & 0.888 & 1.070 & 1.181 & 1.003 & 0.993 & 1.000 & 0.976 & 1.001 & 1.002 & 0.995 & 1.001 & 0.853\\
		\hline
		100 & 1.000 & 0.908 & 1.056 & 1.150 & 1.003 & 0.994 & 1.000 & 0.980 & 1.001 & 1.002 & 0.996 & 1.001 & 0.881\\
		\hline
		120 & 1.000 & 0.924 & 1.046 & 1.126 & 1.002 & 0.996 & 1.000 & 0.985 & 1.001 & 1.001 & 0.997 & 1.001 & 0.911\\
		\hline
		140 & 1.000 & 0.939 & 1.039 & 1.108 & 1.002 & 0.997 & 1.000 & 0.989 & 1.001 & 1.001 & 0.998 & 1.000 & 0.944\\
\hline
	\end{tabular}
\end{table}
\clearpage
\begin{table}[pos=h]
	\renewcommand{\arraystretch}{1.5}
	\begin{tabular}{|c||c|c|c|c|c|c|c|c|c|c|c|c|c|}
		\hline
		$T~(K)$ & $^{113}$In & $^{115}$In & $^{121}$Sb & $^{123}$Te & $^{124}$Xe & $^{135}$Xe & $^{132}$Ba & $^{133}$Cs & $^{138}$Ce & $^{143}$Nd & $^{148}$Pm & $^{149}$Sm & $^{151}$Sm\\
		\hline
		\hline
		20 & 0.980 & 0.969 & 0.994 & 0.981 & 0.993 & 0.691 & 1.000 & 0.996 & 1.003 & 1.007 & 1.000 & 0.628 & 1.348\\
		\hline
		40 & 0.982 & 0.973 & 0.995 & 0.983 & 0.994 & 0.720 & 1.000 & 0.997 & 1.003 & 1.006 & 1.000 & 0.662 & 1.303\\
		\hline
		60 & 0.984 & 0.976 & 0.996 & 0.985 & 0.995 & 0.753 & 1.000 & 0.997 & 1.002 & 1.006 & 1.000 & 0.701 & 1.260\\
		\hline
		80 & 0.987 & 0.980 & 0.996 & 0.988 & 0.996 & 0.789 & 1.000 & 0.998 & 1.002 & 1.005 & 1.000 & 0.746 & 1.221\\
		\hline
		100 & 0.989 & 0.983 & 0.997 & 0.990 & 0.996 & 0.827 & 1.000 & 0.998 & 1.002 & 1.004 & 1.000 & 0.801 & 1.185\\
		\hline
		120 & 0.992 & 0.987 & 0.998 & 0.992 & 0.997 & 0.868 & 1.000 & 0.998 & 1.001 & 1.003 & 1.000 & 0.866 & 1.151\\
		\hline
		140 & 0.994 & 0.991 & 0.998 & 0.994 & 0.998 & 0.909 & 1.000 & 0.999 & 1.001 & 1.002 & 1.000 & 0.942 & 1.119\\
\hline
	\end{tabular}
\end{table}
\clearpage
\begin{table}[pos=h]
	\renewcommand{\arraystretch}{1.5}
	\begin{tabular}{|c||c|c|c|c|c|c|c|c|c|c|c|c|c|}
		\hline
		$T~(K)$ & $^{152}$Sm & $^{151}$Eu & $^{152}$Eu & $^{153}$Eu & $^{154}$Eu & $^{155}$Eu & $^{152}$Gd & $^{155}$Gd & $^{157}$Gd & $^{159}$Tb & $^{156}$Dy & $^{158}$Dy & $^{160}$Dy\\
		\hline
		\hline
		20 & 0.994 & 1.408 & 1.300 & 1.030 & 0.813 & 0.987 & 1.007 & 0.834 & 0.791 & 0.998 & 0.986 & 1.021 & 0.990\\
		\hline
		40 & 0.995 & 1.353 & 1.262 & 1.027 & 0.832 & 0.989 & 1.006 & 0.861 & 0.820 & 0.998 & 0.988 & 1.019 & 0.991\\
		\hline
		60 & 0.996 & 1.303 & 1.226 & 1.024 & 0.853 & 0.990 & 1.006 & 0.883 & 0.846 & 0.999 & 0.989 & 1.016 & 0.992\\
		\hline
		80 & 0.996 & 1.258 & 1.193 & 1.020 & 0.875 & 0.991 & 1.005 & 0.900 & 0.868 & 0.999 & 0.991 & 1.014 & 0.993\\
		\hline
		100 & 0.997 & 1.216 & 1.161 & 1.017 & 0.898 & 0.993 & 1.004 & 0.911 & 0.884 & 0.999 & 0.992 & 1.012 & 0.995\\
		\hline
		120 & 0.998 & 1.177 & 1.131 & 1.013 & 0.923 & 0.995 & 1.003 & 0.917 & 0.896 & 0.999 & 0.994 & 1.009 & 0.996\\
		\hline
		140 & 0.998 & 1.142 & 1.102 & 1.010 & 0.950 & 0.997 & 1.002 & 0.919 & 0.902 & 1.000 & 0.996 & 1.007 & 0.997\\
\hline
	\end{tabular}
\end{table}
\clearpage
\begin{table}[pos=h]
	\renewcommand{\arraystretch}{1.5}
	\begin{tabular}{|c||c|c|c|c|c|c|c|c|c|c|c|c|c|}
		\hline
		$T~(K)$ & $^{161}$Dy & $^{162}$Dy & $^{163}$Dy & $^{164}$Dy & $^{167}$Er & $^{169}$Tm & $^{168}$Yb & $^{175}$Lu & $^{176}$Lu & $^{174}$Hf & $^{176}$Hf & $^{177}$Hf & $^{178}$Hf\\
		\hline
		\hline
		20 & 1.016 & 0.992 & 0.981 & 1.024 & 0.918 & 0.992 & 0.925 & 0.996 & 0.718 & 1.046 & 0.996 & 0.969 & 0.995\\
		\hline
		40 & 1.014 & 0.993 & 0.983 & 1.021 & 0.927 & 0.993 & 0.934 & 0.997 & 0.745 & 1.041 & 0.997 & 0.973 & 0.996\\
		\hline
		60 & 1.012 & 0.994 & 0.985 & 1.019 & 0.936 & 0.994 & 0.942 & 0.997 & 0.775 & 1.035 & 0.997 & 0.976 & 0.996\\
		\hline
		80 & 1.011 & 0.995 & 0.988 & 1.016 & 0.946 & 0.995 & 0.951 & 0.998 & 0.809 & 1.030 & 0.998 & 0.980 & 0.997\\
		\hline
		100 & 1.009 & 0.996 & 0.990 & 1.013 & 0.955 & 0.996 & 0.959 & 0.998 & 0.847 & 1.025 & 0.998 & 0.983 & 0.997\\
		\hline
		120 & 1.007 & 0.997 & 0.992 & 1.010 & 0.965 & 0.997 & 0.968 & 0.999 & 0.891 & 1.020 & 0.998 & 0.987 & 0.998\\
		\hline
		140 & 1.005 & 0.997 & 0.994 & 1.008 & 0.975 & 0.998 & 0.978 & 0.999 & 0.944 & 1.015 & 0.999 & 0.991 & 0.998\\
\hline
	\end{tabular}
\end{table}
\clearpage
\begin{table}[pos=h]
	\renewcommand{\arraystretch}{1.5}
	\begin{tabular}{|c||c|c|c|c|c|c|c|c|c|c|c|c|}
		\hline
		$T~(K)$ & $^{179}$Hf & $^{180}$Hf & $^{180}$Ta & $^{181}$Ta & $^{182}$Ta & $^{180}$W & $^{182}$W & $^{185}$Re & $^{187}$Re & $^{186}$Os & $^{187}$Os & $^{191}$Ir\\
		\hline
		\hline
		20 & 1.005 & 1.001 & 0.965 & 0.994 & 0.728 & 1.005 & 0.995 & 0.992 & 1.011 & 1.040 & 1.032 & 1.019\\
		\hline
		40 & 1.004 & 1.001 & 0.961 & 0.995 & 0.754 & 1.005 & 0.995 & 0.993 & 1.010 & 1.035 & 1.028 & 1.017\\
		\hline
		60 & 1.004 & 1.001 & 0.948 & 0.996 & 0.783 & 1.004 & 0.996 & 0.994 & 1.008 & 1.031 & 1.025 & 1.015\\
		\hline
		80 & 1.003 & 1.001 & 0.943 & 0.996 & 0.816 & 1.004 & 0.997 & 0.995 & 1.007 & 1.026 & 1.021 & 1.012\\
		\hline
		100 & 1.003 & 1.001 & 0.944 & 0.997 & 0.852 & 1.003 & 0.997 & 0.996 & 1.006 & 1.022 & 1.018 & 1.010\\
		\hline
		120 & 1.002 & 1.000 & 0.948 & 0.998 & 0.894 & 1.002 & 0.998 & 0.997 & 1.005 & 1.017 & 1.014 & 1.008\\
		\hline
		140 & 1.002 & 1.000 & 0.953 & 0.998 & 0.942 & 1.002 & 0.998 & 0.997 & 1.004 & 1.013 & 1.011 & 1.007\\
\hline
	\end{tabular}
\end{table}
\clearpage
\begin{table}[pos=h]
	\renewcommand{\arraystretch}{1.5}
	\begin{tabular}{|c||c|c|c|c|c|c|c|c|c|c|c|c|}
		\hline
		$T~(K)$ & $^{192}$Ir & $^{193}$Ir & $^{190}$Os & $^{196}$Hg & $^{199}$Hg & $^{197}$Au & $^{204}$Hg & $^{229}$Th & $^{232}$Th & $^{231}$Pa & $^{233}$Pa & $^{234}$U\\
		\hline
		\hline
		20 & 0.860 & 0.972 & 1.023 & 1.024 & 1.021 & 0.992 & 0.362 & 0.955 & 1.011 & 1.210 & 1.112 & 1.019\\
		\hline
		40 & 0.875 & 0.975 & 1.021 & 1.021 & 1.019 & 0.993 & 0.481 & 0.960 & 1.010 & 1.180 & 1.098 & 1.017\\
		\hline
		60 & 0.890 & 0.979 & 1.018 & 1.018 & 1.017 & 0.994 & 0.568 & 0.965 & 1.009 & 1.153 & 1.084 & 1.015\\
		\hline
		80 & 0.907 & 0.982 & 1.015 & 1.016 & 1.014 & 0.995 & 0.640 & 0.970 & 1.007 & 1.130 & 1.072 & 1.013\\
		\hline
		100 & 0.924 & 0.985 & 1.013 & 1.013 & 1.012 & 0.996 & 0.702 & 0.976 & 1.006 & 1.109 & 1.060 & 1.010\\
		\hline
		120 & 0.942 & 0.988 & 1.010 & 1.010 & 1.009 & 0.997 & 0.757 & 0.981 & 1.005 & 1.091 & 1.049 & 1.008\\
		\hline
		140 & 0.961 & 0.991 & 1.008 & 1.008 & 1.007 & 0.998 & 0.808 & 0.987 & 1.004 & 1.075 & 1.038 & 1.006\\
\hline
	\end{tabular}
\end{table}
\clearpage
\begin{table}[pos=h]
	\renewcommand{\arraystretch}{1.5}
	\begin{tabular}{|c||c|c|c|c|c|c|c|c|c|c|c|c|}
		\hline
		$T~(K)$ & $^{235}$U & $^{238}$U & $^{237}$Np & $^{239}$Pu & $^{240}$Pu & $^{241}$Pu & $^{241}$Am & $^{242}$Am & $^{243}$Am & $^{249}$Cf & $^{252}$Cf & $^{249}$Bk\\
		\hline
		\hline
		20 & 1.094 & 0.998 & 1.046 & 0.959 & 0.958 & 1.197 & 1.125 & 0.933 & 0.980 & 1.130 & 1.010 & 0.854\\
		\hline
		40 & 1.091 & 0.998 & 1.040 & 0.956 & 0.962 & 1.169 & 1.106 & 0.940 & 0.983 & 1.114 & 1.009 & 0.868\\
		\hline
		60 & 1.084 & 0.998 & 1.035 & 0.957 & 0.967 & 1.144 & 1.091 & 0.948 & 0.985 & 1.098 & 1.008 & 0.884\\
		\hline
		80 & 1.074 & 0.999 & 1.030 & 0.962 & 0.972 & 1.122 & 1.077 & 0.955 & 0.987 & 1.084 & 1.007 & 0.901\\
		\hline
		100 & 1.063 & 0.999 & 1.025 & 0.970 & 0.977 & 1.103 & 1.064 & 0.963 & 0.989 & 1.070 & 1.006 & 0.920\\
		\hline
		120 & 1.053 & 0.999 & 1.020 & 0.980 & 0.982 & 1.086 & 1.054 & 0.971 & 0.992 & 1.057 & 1.005 & 0.943\\
		\hline
		140 & 1.043 & 0.999 & 1.015 & 0.992 & 0.987 & 1.072 & 1.044 & 0.979 & 0.994 & 1.044 & 1.003 & 0.968\\
\hline
	\end{tabular}
\end{table}

\clearpage

\begin{table}[pos=ht!]
	\centering{}\caption{Ideal guide-distribution Westcott $g_T$ factors at various temperatures $T$ for non-$1/v$ nuclei listed in References~\citep{IAEA, Molnar, vanSluijs15, Holden99, Pritychenko25}, calculated using the cross-section integration method in the \texttt{WestcottFactors} toolkit~\citep{WestcottFactors-GitHub, WestcottFactors-PyPI}, with cross sections taken from ENDF/B-VIII.1~\citep{Brown18,Nobre24}. \protect\label{Table:Ideal_guide_g-factors}}
	\renewcommand{\arraystretch}{1.5}
\begin{tabular}{|c||c|c|c|c|c|c|c|c|c|c|c|c|c|}
		\hline
		$T~(K)$ & $^{30}$Si & $^{36}$S & $^{36}$Ar & $^{38}$Ar & $^{83}$Kr & $^{87}$Sr & $^{93}$Nb & $^{103}$Rh & $^{105}$Pd & $^{107}$Ag & $^{109}$Ag & $^{111}$Cd & $^{113}$Cd\\
		\hline
		\hline
		20 & 0.959 & 0.636 & 0.998 & 1.033 & 0.964 & 0.949 & 0.959 & 0.922 & 0.961 & 0.962 & 0.951 & 0.960 & 0.737\\
		\hline
		40 & 0.971 & 0.685 & 1.038 & 1.102 & 0.976 & 0.961 & 0.971 & 0.935 & 0.973 & 0.974 & 0.963 & 0.972 & 0.753\\
		\hline
		60 & 0.977 & 0.715 & 1.050 & 1.129 & 0.981 & 0.966 & 0.976 & 0.942 & 0.978 & 0.979 & 0.969 & 0.978 & 0.765\\
		\hline
		80 & 0.980 & 0.736 & 1.054 & 1.138 & 0.984 & 0.970 & 0.980 & 0.946 & 0.981 & 0.982 & 0.972 & 0.981 & 0.775\\
		\hline
		100 & 0.982 & 0.753 & 1.055 & 1.141 & 0.986 & 0.973 & 0.982 & 0.949 & 0.983 & 0.984 & 0.974 & 0.983 & 0.784\\
		\hline
		120 & 0.984 & 0.767 & 1.055 & 1.141 & 0.987 & 0.974 & 0.983 & 0.952 & 0.985 & 0.986 & 0.976 & 0.984 & 0.794\\
		\hline
		140 & 0.985 & 0.780 & 1.054 & 1.139 & 0.988 & 0.976 & 0.985 & 0.955 & 0.986 & 0.987 & 0.978 & 0.986 & 0.804\\
\hline
	\end{tabular}
\end{table}
\clearpage
\begin{table}[pos=h]
	\renewcommand{\arraystretch}{1.5}
	\begin{tabular}{|c||c|c|c|c|c|c|c|c|c|c|c|c|c|}
		\hline
		$T~(K)$ & $^{113}$In & $^{115}$In & $^{121}$Sb & $^{123}$Te & $^{124}$Xe & $^{135}$Xe & $^{132}$Ba & $^{133}$Cs & $^{138}$Ce & $^{143}$Nd & $^{148}$Pm & $^{149}$Sm & $^{151}$Sm\\
		\hline
		\hline
		20 & 0.938 & 0.928 & 0.953 & 0.939 & 0.952 & 0.647 & 0.959 & 0.955 & 0.962 & 0.966 & 0.959 & 0.584 & 1.324\\
		\hline
		40 & 0.951 & 0.940 & 0.965 & 0.952 & 0.964 & 0.664 & 0.971 & 0.967 & 0.974 & 0.978 & 0.971 & 0.602 & 1.325\\
		\hline
		60 & 0.957 & 0.947 & 0.971 & 0.958 & 0.970 & 0.677 & 0.977 & 0.973 & 0.979 & 0.983 & 0.977 & 0.616 & 1.317\\
		\hline
		80 & 0.961 & 0.951 & 0.974 & 0.962 & 0.973 & 0.690 & 0.980 & 0.976 & 0.982 & 0.986 & 0.980 & 0.630 & 1.307\\
		\hline
		100 & 0.964 & 0.954 & 0.976 & 0.964 & 0.976 & 0.702 & 0.982 & 0.978 & 0.984 & 0.988 & 0.982 & 0.645 & 1.296\\
		\hline
		120 & 0.966 & 0.957 & 0.978 & 0.967 & 0.977 & 0.715 & 0.983 & 0.980 & 0.986 & 0.989 & 0.983 & 0.662 & 1.285\\
		\hline
		140 & 0.968 & 0.959 & 0.980 & 0.969 & 0.979 & 0.728 & 0.985 & 0.981 & 0.987 & 0.990 & 0.985 & 0.679 & 1.273\\
\hline
	\end{tabular}
\end{table}
\clearpage
\begin{table}[pos=h]
	\renewcommand{\arraystretch}{1.5}
	\begin{tabular}{|c||c|c|c|c|c|c|c|c|c|c|c|c|c|}
		\hline
		$T~(K)$ & $^{152}$Sm & $^{151}$Eu & $^{152}$Eu & $^{153}$Eu & $^{154}$Eu & $^{155}$Eu & $^{152}$Gd & $^{155}$Gd & $^{157}$Gd & $^{159}$Tb & $^{156}$Dy & $^{158}$Dy & $^{160}$Dy\\
		\hline
		\hline
		20 & 0.953 & 1.387 & 1.272 & 0.990 & 0.768 & 0.946 & 0.966 & 0.782 & 0.739 & 0.957 & 0.945 & 0.981 & 0.949\\
		\hline
		40 & 0.965 & 1.385 & 1.275 & 1.002 & 0.784 & 0.959 & 0.978 & 0.801 & 0.758 & 0.969 & 0.957 & 0.992 & 0.961\\
		\hline
		60 & 0.971 & 1.375 & 1.269 & 1.006 & 0.795 & 0.964 & 0.983 & 0.814 & 0.772 & 0.975 & 0.963 & 0.997 & 0.967\\
		\hline
		80 & 0.974 & 1.362 & 1.261 & 1.008 & 0.804 & 0.968 & 0.986 & 0.824 & 0.783 & 0.978 & 0.967 & 0.999 & 0.970\\
		\hline
		100 & 0.977 & 1.349 & 1.252 & 1.009 & 0.812 & 0.970 & 0.988 & 0.833 & 0.793 & 0.980 & 0.969 & 1.001 & 0.973\\
		\hline
		120 & 0.978 & 1.336 & 1.243 & 1.010 & 0.821 & 0.972 & 0.990 & 0.840 & 0.802 & 0.982 & 0.971 & 1.002 & 0.975\\
		\hline
		140 & 0.980 & 1.323 & 1.234 & 1.010 & 0.830 & 0.974 & 0.991 & 0.846 & 0.809 & 0.983 & 0.973 & 1.002 & 0.976\\
\hline
	\end{tabular}
\end{table}
\clearpage
\begin{table}[pos=h]
	\renewcommand{\arraystretch}{1.5}
	\begin{tabular}{|c||c|c|c|c|c|c|c|c|c|c|c|c|c|}
		\hline
		$T~(K)$ & $^{161}$Dy & $^{162}$Dy & $^{163}$Dy & $^{164}$Dy & $^{167}$Er & $^{169}$Tm & $^{168}$Yb & $^{175}$Lu & $^{176}$Lu & $^{174}$Hf & $^{176}$Hf & $^{177}$Hf & $^{178}$Hf\\
		\hline
		\hline
		20 & 0.975 & 0.951 & 0.939 & 0.984 & 0.875 & 0.950 & 0.882 & 0.955 & 0.673 & 1.006 & 0.955 & 0.927 & 0.954\\
		\hline
		40 & 0.987 & 0.963 & 0.952 & 0.995 & 0.889 & 0.963 & 0.896 & 0.967 & 0.690 & 1.017 & 0.967 & 0.940 & 0.966\\
		\hline
		60 & 0.992 & 0.968 & 0.958 & 1.000 & 0.897 & 0.968 & 0.904 & 0.973 & 0.702 & 1.021 & 0.973 & 0.946 & 0.972\\
		\hline
		80 & 0.994 & 0.972 & 0.962 & 1.002 & 0.903 & 0.972 & 0.910 & 0.976 & 0.714 & 1.023 & 0.976 & 0.951 & 0.975\\
		\hline
		100 & 0.996 & 0.974 & 0.965 & 1.003 & 0.908 & 0.974 & 0.914 & 0.978 & 0.726 & 1.023 & 0.978 & 0.954 & 0.977\\
		\hline
		120 & 0.997 & 0.976 & 0.967 & 1.004 & 0.912 & 0.976 & 0.919 & 0.980 & 0.739 & 1.023 & 0.980 & 0.957 & 0.979\\
		\hline
		140 & 0.998 & 0.978 & 0.969 & 1.005 & 0.916 & 0.978 & 0.923 & 0.981 & 0.752 & 1.023 & 0.981 & 0.959 & 0.980\\
\hline
	\end{tabular}
\end{table}
\clearpage
\begin{table}[pos=h]
	\renewcommand{\arraystretch}{1.5}
	\begin{tabular}{|c||c|c|c|c|c|c|c|c|c|c|c|c|}
		\hline
		$T~(K)$ & $^{179}$Hf & $^{180}$Hf & $^{180}$Ta & $^{181}$Ta & $^{182}$Ta & $^{180}$W & $^{182}$W & $^{185}$Re & $^{187}$Re & $^{186}$Os & $^{187}$Os & $^{191}$Ir\\
		\hline
		\hline
		20 & 0.964 & 0.960 & 0.919 & 0.953 & 0.683 & 0.965 & 0.954 & 0.951 & 0.970 & 1.000 & 0.992 & 0.979\\
		\hline
		40 & 0.976 & 0.972 & 0.933 & 0.965 & 0.699 & 0.976 & 0.966 & 0.963 & 0.982 & 1.011 & 1.003 & 0.990\\
		\hline
		60 & 0.981 & 0.977 & 0.937 & 0.971 & 0.712 & 0.982 & 0.971 & 0.969 & 0.987 & 1.015 & 1.008 & 0.995\\
		\hline
		80 & 0.984 & 0.980 & 0.938 & 0.974 & 0.724 & 0.985 & 0.975 & 0.972 & 0.990 & 1.017 & 1.010 & 0.997\\
		\hline
		100 & 0.986 & 0.982 & 0.938 & 0.977 & 0.735 & 0.987 & 0.977 & 0.975 & 0.991 & 1.018 & 1.011 & 0.999\\
		\hline
		120 & 0.988 & 0.984 & 0.939 & 0.978 & 0.747 & 0.988 & 0.979 & 0.977 & 0.993 & 1.018 & 1.011 & 1.000\\
		\hline
		140 & 0.989 & 0.985 & 0.940 & 0.980 & 0.760 & 0.989 & 0.980 & 0.978 & 0.993 & 1.018 & 1.011 & 1.000\\
\hline
	\end{tabular}
\end{table}
\clearpage
\begin{table}[pos=h]
	\renewcommand{\arraystretch}{1.5}
	\begin{tabular}{|c||c|c|c|c|c|c|c|c|c|c|c|c|}
		\hline
		$T~(K)$ & $^{192}$Ir & $^{193}$Ir & $^{190}$Os & $^{196}$Hg & $^{199}$Hg & $^{197}$Au & $^{204}$Hg & $^{229}$Th & $^{232}$Th & $^{231}$Pa & $^{233}$Pa & $^{234}$U\\
		\hline
		\hline
		20 & 0.816 & 0.931 & 0.983 & 0.984 & 0.981 & 0.951 & 0.199 & 0.913 & 0.970 & 1.182 & 1.076 & 0.979\\
		\hline
		40 & 0.831 & 0.943 & 0.995 & 0.995 & 0.993 & 0.963 & 0.264 & 0.926 & 0.982 & 1.186 & 1.084 & 0.990\\
		\hline
		60 & 0.841 & 0.950 & 0.999 & 1.000 & 0.997 & 0.968 & 0.312 & 0.933 & 0.987 & 1.183 & 1.086 & 0.995\\
		\hline
		80 & 0.848 & 0.954 & 1.002 & 1.002 & 1.000 & 0.972 & 0.351 & 0.938 & 0.990 & 1.177 & 1.085 & 0.998\\
		\hline
		100 & 0.855 & 0.957 & 1.003 & 1.003 & 1.001 & 0.974 & 0.385 & 0.941 & 0.992 & 1.171 & 1.083 & 0.999\\
		\hline
		120 & 0.862 & 0.959 & 1.004 & 1.004 & 1.002 & 0.976 & 0.415 & 0.945 & 0.993 & 1.165 & 1.080 & 1.000\\
		\hline
		140 & 0.869 & 0.962 & 1.004 & 1.004 & 1.002 & 0.978 & 0.442 & 0.947 & 0.994 & 1.159 & 1.077 & 1.001\\
\hline
	\end{tabular}
\end{table}
\clearpage
\begin{table}[pos=h]
	\renewcommand{\arraystretch}{1.5}
	\begin{tabular}{|c||c|c|c|c|c|c|c|c|c|c|c|c|}
		\hline
		$T~(K)$ & $^{235}$U & $^{238}$U & $^{237}$Np & $^{239}$Pu & $^{240}$Pu & $^{241}$Pu & $^{241}$Am & $^{242}$Am & $^{243}$Am & $^{249}$Cf & $^{252}$Cf & $^{249}$Bk\\
		\hline
		\hline
		20 & 1.048 & 0.956 & 1.007 & 0.924 & 0.916 & 1.168 & 1.092 & 0.890 & 0.939 & 1.094 & 0.969 & 0.812\\
		\hline
		40 & 1.061 & 0.969 & 1.017 & 0.934 & 0.929 & 1.172 & 1.099 & 0.903 & 0.951 & 1.103 & 0.981 & 0.826\\
		\hline
		60 & 1.066 & 0.974 & 1.021 & 0.938 & 0.935 & 1.170 & 1.099 & 0.911 & 0.957 & 1.103 & 0.986 & 0.835\\
		\hline
		80 & 1.068 & 0.977 & 1.023 & 0.941 & 0.940 & 1.165 & 1.097 & 0.916 & 0.961 & 1.102 & 0.989 & 0.843\\
		\hline
		100 & 1.068 & 0.980 & 1.023 & 0.944 & 0.943 & 1.159 & 1.094 & 0.921 & 0.964 & 1.099 & 0.991 & 0.850\\
		\hline
		120 & 1.067 & 0.981 & 1.023 & 0.946 & 0.947 & 1.154 & 1.091 & 0.925 & 0.966 & 1.096 & 0.992 & 0.857\\
		\hline
		140 & 1.066 & 0.983 & 1.023 & 0.949 & 0.949 & 1.148 & 1.088 & 0.929 & 0.968 & 1.092 & 0.993 & 0.864\\
\hline
	\end{tabular}
\end{table}

\end{document}